\documentclass[unnumsec,webpdf,contemporary,large]{oup-authoring-template}%

\onecolumn 

\graphicspath{{Fig/}}

\usepackage{cleveref}
\usepackage{graphicx}
\usepackage{subcaption}

\theoremstyle{thmstyleone}%
\theoremstyle{thmstyletwo}%
\theoremstyle{thmstylethree}%

\begin{document}

\journaltitle{Journal Title Here}
\DOI{DOI HERE}
\copyrightyear{2022}
\pubyear{2019}
\access{Advance Access Publication Date: Day Month Year}
\appnotes{Paper}

\firstpage{1}


\title[Heterogeneous Clinical Trial Outcomes via Multi-Output Gaussian Processes]{Heterogeneous Clinical Trial Outcomes via Multi-Output Gaussian Processes}

\author[1,$\ast$]{Owen Thomas}
\author[2]{Leiv Rønneberg}

\authormark{Thomas et al.}

\address[1]{\orgdiv{HØKH}, \orgname{Akershus University Hospital}, \orgaddress{\state{Lørenskog}, \country{Norway}}}
\address[2]{\orgdiv{MRC Biostatistics Unit}, \orgname{University of Cambridge}, \orgaddress{\state{Cambridge}, \country{UK}}}

\corresp[$\ast$]{Corresponding author. \href{email:owen.thomas@ahus.no}{owen.thomas@ahus.no}}

\received{Date}{0}{Year}
\revised{Date}{0}{Year}
\accepted{Date}{0}{Year}



\abstract{Repeated measures, commonly performed in clinical research, facilitate computational acceleration for nonlinear Bayesian nonparametric models and enable exact sampling for non-conjugate inference, when combinations of continuous and discrete endpoints are observed. We make use of Kronecker structure for scaling Gaussian Process models to large-scale, heterogeneous, clinical data sets. Model inference is performed in Stan, and comparisons are made with brms on simulated data and two real clinical data sets, following a radiological image quality theme. Scalable Gaussian Process models compare favourably with parametric models on real data sets with 17,460 observations. Different GP model specifications are explored, with components analogous to random effects, and their theoretical properties are described.}


\maketitle


\section{Introduction}\label{sec1}
Clinical research is often performed with structured data built into the study design, sometimes by repeated measurements of individuals at different time points or locations, or using different measurements methods simultaneously on the same individuals. For example, longitudinal studies follow the same cohort repeatedly at different points using the same measurement process, while many Randomised Control Trials (RCTs) will simultaneously measure primary and secondary endpoints reflecting different aspects of a clinical process.

In the article, we observe that either of these types of structured measurement can be used to enable computational tractability of a class of complex statistical models that would otherwise not scale to real clinical data sets of thousands of measurements. Specifically, appropriate repeated measurements enable the pursuit of exact inference for Gaussian Process (GP)\cite{williams2006gaussian} models by representing their covariance matrices as Kronecker products\cite{saatcci2012scalable}. GPs are Bayesian nonparametric models, that are capable of capturing nonlinear covariate dependence, or multi-output correlations\cite{alvarez2012kernels} that would ordinarily be neglected by commonly-used parametric statistical models.

Here we present and run new GP models using repeated measurements in the covariate structure and multiple heterogeneous (mixed continuous and discrete) outputs: exact inference in Stan\cite{stancore} is scaled to tens of thousands of measurements on the processors of a domestic-issue laptop. The models represent the joint covariance between all of the data as Kronecker-structured, and perform Hamiltonian Monte Carlo (HMC) sampling for hyperparameters, missing output values, and on the latent GP space for non-conjugate inference in the presence of heterogeneous outputs. The methods developed here are of use for general application across clinical research, but the real data sets used in this article follow the theme of radiological image quality, in which the covariates consist of patient characteristics, body locations and time, while the outputs represent measurements of image quality, either continuously "objectively" on the Hounsfield scale\cite{hounsfieldunits} of radiodensity, or discretely "subjectively" from expert evaluations.

\boxedtext{
\section*{Main Article Contributions}%
\begin{itemize}
\item The observation that widely-used repeated measurements in clinical research, either in the form of making the same measurement at different points, or making different measurements simultaneously, facilitates the use of Kronecker structure in the covariance matrix of Gaussian Process models and thereby the scaling of exact sampling for Bayesian nonparametric models to large clinical data sets with modest computational resources.
\item An implementation of the relevant Gaussian Process models in Stan\cite{stancore,rstan} (with wrapper in R), allowing for running the inference on multiple real world data sets, and benchmarking various model specificatons such as covariance function choices.
\item An empirical investigation using real-world clinical data sets into the predictive abilities of the Gaussian Process models compared to standard parametric regression models run in brms\cite{brms}, demonstrating the utility of more complex models.
\end{itemize}
}

\section{Methods}\label{sec2}

In this section, we describe the methodological concepts relevant to the models implemented in this article.

\subsection{Repeated Measurements}\label{sec:repeatedmeasures}

The value of repeated measurements is well-established and widely-understood in clinical research. The use of random effects is common when data is drawn from a structured population for which a hierarchical model is appropriate, for example when measuring different individuals repeatedly, or in a multi-centre study. Further structure emerges if data is collected in a systematic way, for example at consistent follow-up times for the entire population, or at multiple, consistent anatomical locations in the body for scans or biopsies. Data can also exhibit repetitions at the outcome level, when different outcomes of interest are measured at the same locations, individuals, and time points: this is common in RCTs with a combination of primary and secondary endpoints.

Within a regression framework, the multiple outcomes can be represented as a matrix Y, and the covariates corresponding to treatments, locations, times, patient characteristics, or anything else that might influence the outcomes, can be represented as a covariate matrix X. Different clinical study designs will imposed different structure on the covariate matrix X. If we consider a design in which $N_1$ individuals are measured at $N_2$ time points, at $N_3$ anatomical locations, with the time points and anatomical locations being identical between individuals, then we can divide columnwise the long-format covariate matrix X of height $N_1 N_2 N_3$ into matrices $X_1$, $X_2$, and $X_3$, with $X_1$ containing information $x_1$ about the $N_1$ unique individuals, repeated $N_2 N_3$ times, $X_2$ containing information $x_2$ about the $N_2$ unique times of measurement, repeated $N_1 N_3$ times, and $X_3$ containing information $x_3$ about the $N_3$ unique anatomical locations, repeated $N_1 N_2$ times. There are in addition $N_4$ distinct outcomes measured for every value of $X$, resulting in an outcome matrix Y of size $N_1 N_2 N_3$ by $N_4$.

\subsection{Gaussian Processes}

Gaussian Processes (GPs) are nonlinear, Bayesian models designed for flexible, probabilistic supervised learning \cite{williams2006gaussian}. They model a dependent variable $y$ conditional on independent variables $X$, via a mean function $m(x)$ and covariance\footnote{also known as kernel function} function $k(x,x',\theta)$, defining a latent variable $f$ that is joint-normally distributed over all the observed data points. The covariance functions are described by hyperparameters $\theta$ that can be learned from data, while the mean function will be taken to be zero here with no loss of generality. The latent function $f$ can be passed through a Gaussian likelihood with a noise variance $\sigma^2$ to model a continuous output $y$:

\begin{align}
f \sim & \mathcal{GP}(0, k(x,x',\theta))\\
y \vert x = & f(x) + \epsilon, \quad \epsilon \sim \mathcal{N}(0, \sigma^2) \nonumber
\end{align}

Specific choices of covariance functions $k(x,x',\theta)$ have corresponding implicit parametric basis functions. One advantage of the "function-space" formulation is the ability to use a finite representation of a function with a potentially infinite-dimensional parametric representation: this is the sense in which the models are considered "nonparametric", in that they avoid specifying a parametric model for the latent mean function.

\subsection{Scalability and Kronecker-Structure Covariance Matrices}

Asserting a GP with a covariance function $k(x,x',\theta)$ over a data set with $N$ covariate observations defines a $N\times N$ covariance\footnote{also known as kernel matrix} matrix $ \mathbf{K}$, where the matrix element $\mathbf{K}[i,j]$ is equal to the covariance function $k(x^i,{x^j}',\theta)$ evaluated at the $ith$ and $jth$ data point. One challenge of working with GPs is the need to perform a decomposition of the $N\times N$ covariance matrix, which is the dominant computational demand when evaluating the marginal likelihood for sampling or computing the predictive distributions. This results in $\mathcal{O}(N^3)$ cubic scaling in computational costs with the number of data points $N$, ruling out exact inference for generic covariance matrices for larger data sets . Various methods exist to enable \textit{approximate inference} for larger data sets\cite{liu2020gaussian,thomas2017scalable}, while exact inference is possible for larger data sets when there is structure in the covariance matrix that can be exploited, such as Kronecker or Toeplitz\cite{saatcci2012scalable}. In instances where data can be represented as lying in a grid structure, with separable covariance structure between each dimension of the grid, the full covariance matrix can be represented as a Kronecker product between the dimensions. A Kronecker product is an operation performed on two matrices generating a third matrix composed of each individual element of the first matrix separately multiplied with the entire second matrix, combined in a blockwise fashion, i.e. for two matrices $A$ and $B$:

\begin{equation}
\mathbf{A} \otimes \mathbf{B} = \begin{bmatrix}
a_{11}\mathbf{B} & \cdots & a_{1n}\mathbf{B}\\
\vdots & \ddots & \vdots \\
a_{n1}\mathbf{B} & \cdots & a_{nn}\mathbf{B}
\end{bmatrix}	
\end{equation}

For a data set consisting of covariates measured over a two dimensional grid, for example the pixels of an image, the covariance matrix over the entire data set ($\mathbf{K}_{12} = K([x_1,x_2],[x'_1,x'_2])$) can be represented as the Kronecker product between two covariance matrices representing each grid dimension independently ($\mathbf{K}_{1} = K(x_1,x'_1)$, $\mathbf{K}_{2} = K(x_2,x'_2)$),  with the constraint that the covariance functions used are separable, i.e. for a grid composed of $n_1$ vertical grid points $x_1$, and $n_2$ horizontal grid points $x_2$, then:

\begin{align}
 \mathbf{K}_{12} = \mathbf{K}_{1} \otimes \mathbf{K}_{2}.
\end{align}

We can apply this reasoning to clinical scenario described in Section \ref{sec:repeatedmeasures}: for a longitudinal study consisting of $N_1$ individuals measured at $N_2$ time points at $N_3$ anatomical locations, the covariate data X can be represented as lying on a three dimensional grid, where one of the grid dimensions $x_1$ is the variation between individuals, the second $x_2$ as the variation in time, and the third $x_3$ variation in anatomical location. Consequently, defining separate covariance matrices representing the variation across individuals (the $N_1\times N_1$ matrix $\mathbf{K}_{1}$), variation over time (the $N_2\times N_1$ matrix $\mathbf{K}_{2}$), and anatomical variation (the $N_3\times N_3$ matrix $\mathbf{K}_{3}$) we can represent the covariance between the covariate measurements $X$ as:

\begin{equation}
    \mathbf{K}_{X} = \mathbf{K}_1  \otimes \mathbf{K}_2  \otimes \mathbf{K}_3.
\end{equation}

Conveniently, when a decomposition of the full matrix is required, the components of the Kronecker product can be decomposed separately, such that the $\mathcal{O}(N^3) = \mathcal{O}(N_1^3 N_2^3 N_3^3)$ computational demands become $\mathcal{O}(N_1^3 + N_2^3 + N_3^3)$. For datasets with appropriate grid structure or repeated measurements, this enables exact inference for tens of thousands of data points on a personal computer. This has been used previously for image data, where the pixels lie on a regularly spaced grid, or when there are multiple endpoints evaluated at the same covariate locations. In this article, we note that the widespread use of repeated measurements in clinical research, can be represented within the Kronecker structure described above.

\subsection{Heterogeneous Multi-Task Gaussian Processes}

The GP framework extends naturally to multi-dimensional responses $\mathbf{Y}$, analogous to the case of multivariate regression\cite{alvarez2012kernels,bonilla2007multi}. The different response dimensions (also known as "tasks" or "outputs") are appended into a single vector, and a between-task covariance matrix $\mathbf{K}_{out}$ is used to model correlations between the tasks. When the different tasks are evaluated at the same values of the covariates, generating a covariance matrix between covariates $\mathbf{K}_{X}$ then Kronecker structure again emerges in the full covariance matrix $\mathbf{K}_{f}= \mathbf{K}_{out}\otimes \mathbf{K}_{X}$, and computational accelerations become possible.

Different tasks often correspond to different methods of evaluating some outcome, and  often exhibit heterogeneity of distributions, e.g. each task may be variously continuous valued, binary or follow some other distribution. In this case, each output will require different likelihood functions to map from the latent function $f$ to observation $y$. For likelihoods other than Gaussian, the latent variable cannot be integrated out analytically and inference must be performed for the latent variable, through sampling, variational inference, or another approximation. 

All of the data sets used in this article use one Kronecker component for the multi-output correlation, and three Kronecker grid components for the covariates, which are grouped into grid dimensions $x_1$, $x_2$ and $x_3$, which may represent variously patient characteristics, intervention treatment, time, body level, or similar. The full covariance structure is therefore:

\begin{equation}
\mathbf{K}_{f} = \mathbf{K}_{out} \otimes \mathbf{K}_1  \otimes \mathbf{K}_2  \otimes \mathbf{K}_3
\end{equation}

\subsection{Covariance Matrix Design and Random Effects}\label{seq:randomeffects}

Imposing Kronecker structure on the full covariance matrix puts some constraints on the types of covariance functions that can be used. Principally, the covariance function must be separable between the different grid components $x_1$ and $x_2$, meaning the overall covariance function can be represented as the product of covariance functions defined over of the grid components, i.e. $k([x_1,x_2],[x'_1,x'_2]) = k_1(x_1,x'_1)k_2(x_2,x'_2)$. Many commonly used multi-dimensional covariance functions exhibit this property, but some designs that might be desirable for interpretation do not exhibit separability. Two are discussed below: additive covariance functions and random effects covariance functions.

An additive covariance function represents a common covariance function over different dimensions as a sum of separate covariance functions\cite{duvenaud2014automatic}, i.e. $k([x_1,x_2],[x'_1,x'_2]) = k_1(x_1,x'_1) + k_2(x_2,x'_2)$. Additive covariance functions might be desired if we are interested in isolating the particular contribution to the output variation from one particular dimension. This is especially useful in medical research when one intervention or treatment is considered to be of central clinical relevance. Within the Kronecker framework, it is possible to assert an additive covariance function within each component of the Kronecker product, but the resulting covariance function over the entire data set is a product of the sum within the component with the covariance functions over the rest of the dimensions, making it different to interpret the sum components separately, i.e.:

\begin{equation}
(\mathbf{K_{1}} + \mathbf{K_{2}}) \otimes \mathbf{K}_{3} = (\mathbf{K}_{1} \otimes \mathbf{K_{3}})  + (\mathbf{K}_{2} \otimes\mathbf{K_{3}}) 
\end{equation}

Random effects are often desirable in the presence of repeated measurements, where we are not necessarily interested in the variation between individuals but we would still like to include it in the model \cite{gelman2007data}. This can be performed in an elegant way within GP regression models by including a structured diagonal noise component representing the random variation across individuals. In the Kronecker context, this can be achieved by adding spherical noise to the Kronecker product component that represents the variation between individuals that call for additional random effects, representing the "noise" sampled when moving between repeated measurements. As this is an additive covariance function where one of the covariance functions is diagonal noise, the same problem emerges when combining additive covariance functions with Kronecker structure: the resulting covariance function components cannot be interpreted totally straightforwardly, as they are multiplied with the covariance functions associated with the other Kronecker components. Formally, a traditional random effects model is represented in the first half of the following inequality, and the model we implemented in the second:

\begin{equation}\label{eq:randomeffects}
 (\mathbf{K}_{1} \otimes \mathbf{K}_{2}) + (\sigma^2 \mathcal{I}_{1} \otimes \mathcal{I}_{2}) \neq  (\mathbf{K_{1}} + \sigma^2 \mathcal{I}_{1}) \otimes \mathbf{K}_{2} 
\end{equation}

This formulation of a random effect is equivalent to adding some extra variance to the coefficients corresponding to non-patient specific covariates. We detail this in Appendix \ref{app:re-icm}, for the case where all kernels are linear and thus GP regression is equivalent to Bayesian linear regression.

We ran separate models with and without the "random effect" noise in the individual-level covariance function to explore its influence on the model fit and predictions. We call the models with and without this the mixed-effect GP ("GP.m") and the fixed-effect GP ("GP.f").

The covariance functions for each covariate were chosen conditional on the data. For continuous-valued covariates, a squared-exponential covariance function with a lengthscale hyperparameter enabling automatic relevance determination (ARD) was used. For binary covariates, a linear covariance function was used with a variance hyperparameter, as a more complex covariance function would be unnecessary for binary data. Nominal or ordinal covariates were one-hot encoded to corresponding binary variables and a linear covariance function used. The multi-output covariance matrix $\mathbf{K}_{out}$ is parametrised as a Cholesky-decomposed correlation matrix multiplied with a diagonal matrix representing the separate output variances.

\subsection{Full Model Specification}

In summary, the three-component Kronecker model used in this article, with $n_g$ Gaussian-distributed outputs $y_{g}$, $n_b$ Bernoulli-distributed outputs $y_{b}$, and repeated covariates partitioned into repeated measure grid components $x_1$, $x_2$ and $x_3$, becomes:

\begin{center}
\begin{tabular}{ l l }
 $\rho_1, \rho_2, \rho_3 \sim \mathrm{InvGamma}(2,1)$ & Lengthscales for each kernel Kronecker component  \\ 
 $\alpha \sim \mathrm{InvGamma}(2,1)$ & Kernel variances for each output  \\  
 $\alpha^{(n)} \sim \mathrm{InvGamma}(2,1)$ & Noise variance for each Gaussian-distributed output\\
 $\eta \sim \mathcal{N}(0,1)$ & Standard Normal latent $\eta$ for computation purposes \\
$L \sim \mathrm{LkjCholesky}(3)$ & Correlation matrix between outputs with LkjCholesky prior\\
$L^{(n)} \sim \mathrm{LkjCholesky}(3)$ & Noise correlation matrix between Gaussian outputs with LkjCholesky prior\\
$\sigma^2_{re} \sim \mathrm{InvGamma}(2,1)$ & Variance of optional random effects kernel component\\
$K_1 = k(x_1,x_1,\rho_1)$ & First Kronecker kernel matrix component\\
$K_2 = k(x_2,x_2,\rho_2)$ & Second Kronecker kernel matrix component\\
$K_3 = k(x_3,x_3,\rho_3) + \sigma^2_{re} \mathcal{I}$ & Third Kronecker kernel matrix component, with optional random effects\\
$f =  \mathrm{diag}(\alpha) L \otimes \mathrm{chol}(K_3) \otimes \mathrm{chol}(K_2) \otimes \mathrm{chol}(K_1) \eta$ &Latent variable $f$ constructed from kernels and $\eta$\\
$\Sigma^{(n)} =  \mathrm{diag}(\alpha^{(n)}) L^{(n)} ( \mathrm{diag}(\alpha^{(n)}) L^{(n)}) ^ T$& Noise covariance between Gaussian-distributed outputs\\
$y_{g} \sim \mathcal{N}(f[1:n_{g} , ], \Sigma^{(n)} )$& Distribution of $n_g$ Gaussian-distributed outputs $y_{g}$\\
$y_{b} \sim \mathrm{Bernoulli}(\Phi(f[n_{g} + 1:n_{g} + n_{b} , ]))$ &Distribution of $n_b$ Bernoulli-distributed outputs $y_{b}$
\end{tabular}
\end{center}

\subsection{Stan Implementation}

The methods described here were implemented in Stan\cite{stancore}, a probabilistic programming language designed for Bayesian inference, in which model specification is performed explicitly within the language, and Hamiltonian Monte Carlo\cite{hoffman2014no} (HMC) with the No U-Turn Sampler (NUTS) is performed for the model parameters. The Gaussian Process was represented for inference with a standardised, uncorrelated Gaussian latent variable $\eta$, which was reshaped and transformed by the covariance Kronecker components to form the latent variable $f$ conditioned on the covariance. HMC was performed for the GP hyperparameters and noise terms, the latent function representation $\eta$, the latent function $f$ for discrete-valued tasks, and the missing values of the outputs. The Stan programs were called from within the R programming language via Rstan\cite{rstan}. Similar computational speedups have been achieved in Stan previously \cite{flaxman2015fast, Ronneberg2021}.

\subsection{Missing Data}

In the case of missing values of the output variable y, which often arise when exploiting Kronecker structure over an incomplete grid, the Bayesian framework offers a convenient solution: the missing y values are treated as parameters to be inferred, using the distribution implied by the latent variable $f$ and likelihood $p(y|f)$, conditional on the observed covariates. The scalability associated with the Kronecker decomposition is preserved, while interpretable posterior distributions for the unobserved data are provided.

The inference for continuous-valued missing outputs $y$ can be integrated in a straightforward way with most inference schemes. Given that Stan cannot perform inference over discrete variables, missing binary outputs pose a challenge. This was resolved by calculating the log-likelihood contributions according to the Bernoulli likelihood with an appropriate link function, with the missing $y$ varying continuously between zero and one. This allows for efficient gradient-informed inference procedures, and the interpretation of the imputed variables as probabilities.

\subsection{Radiological Image Quality}

While the models described here are of interest across many applications across and beyond clinical research, the application theme of this article is radiological image quality, a field in which it is common to combine continuous-valued "objective" measurements of image quality with discrete-valued "subjective" expert evaluations of image quality. This article uses two real data sets, the first of which is observational time-series data, and the second of which comes from a RCT with thorough repeated measures. In each case the effect of primary interest is the effect on the image quality of volume of contrast medium per body weight. 

The first real data set was previously published as a clinical study in \cite{tran2022indirect}. The image quality was measured in venal blood vessels of 53 patients, with continuous-valued attenuation in Hounsfield Units as the objective endpoint, and binary evaluations of image quality from three different consultant radiologists as the subjective endpoints. Each patient was injected with a different quantity of contrast medium per body weight, and evaluations of the images were taken at six thirty-second interval time points after injection. Evaluations were also performed for each side of the body. Further covariates used were gender, age and tube voltage. The combination of 53 patients, 4 endpoints, 7 time points and 2 body sides implies a covariance structure of size 2,968, but the use of the Kronecker product for each of these contributions means that the computations are readily tractable on a personal laptop.

\begin{figure}
    \centering
    \includegraphics[width=0.9\textwidth]{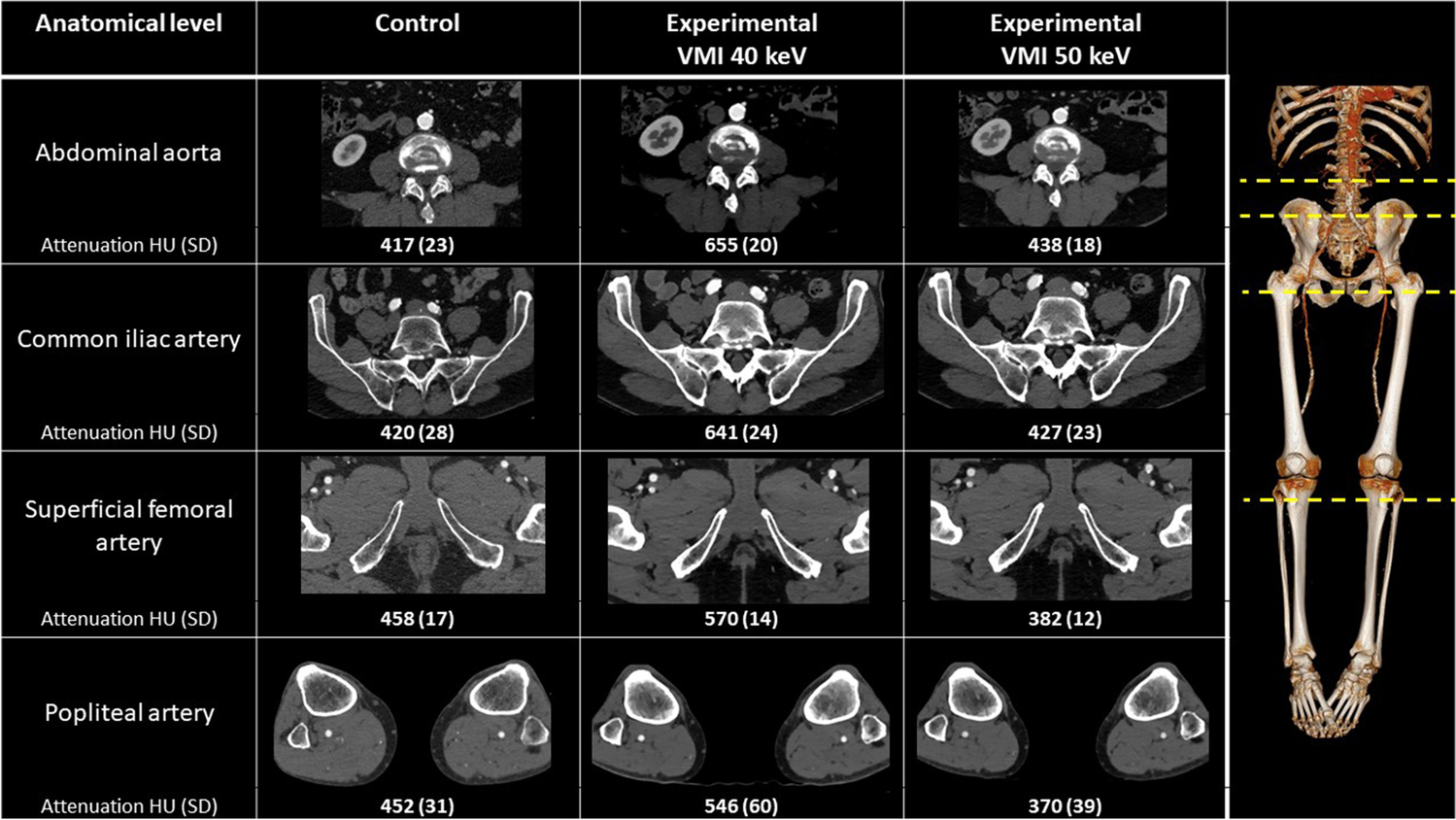}
    \caption{Lower extremity CTA in the control group and in the experimental group with energy levels at 50 and 40 keV. ROIs  were placed in the abdominal aorta, common iliac artery, superficial femoral artery, and popliteal artery show attenuation and noise values. Reproduced with permission from \cite{kristiansen2023halved}.}
    \label{fig:levels-image}
\end{figure}

\begin{figure}
\centering
   \includegraphics[width=0.48\textwidth]{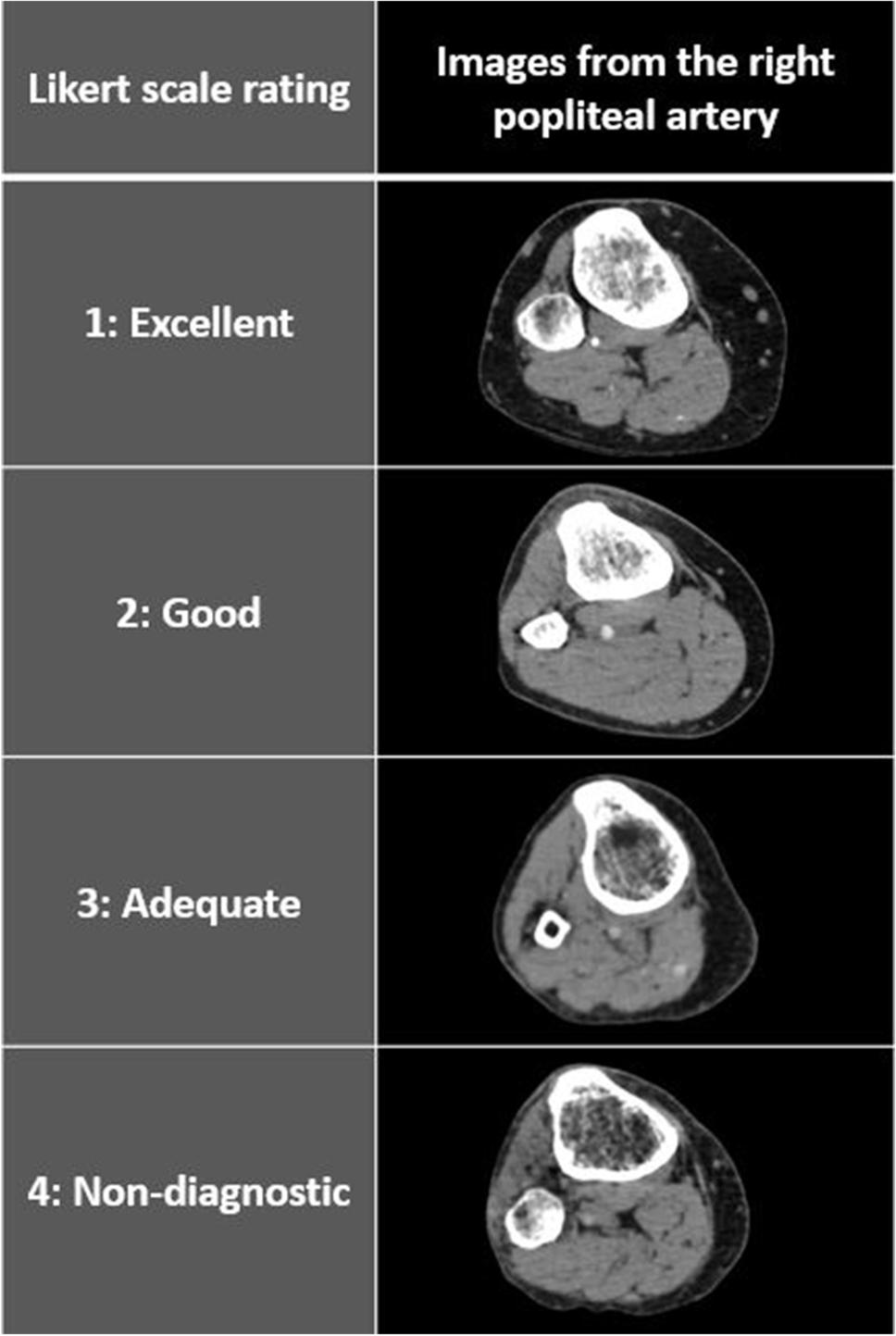}
    \caption{Examples of ordinal scale for subjective examination quality. Reproduced with permission from \cite{kristiansen2023halved}.}
   \label{fig:ordinal-scale}
\end{figure}

The second real data set comes from a Randomised Control Trial assessing image quality in arterial blood vessels following Computed Tomography Angiography (CTA) \cite{kristiansen2023halved}. 210 patients were randomised to receive full versus half doses of contrast volume per body weight, with measurements repeated at 7 different locations in the body, and 3 different spectral energy levels measured in kiloelectronVolts (keV). Examples of images recorded at different body levels and energy levels are presented in Figure \ref{fig:levels-image}. Attenuation level and image noise recorded in selected Regions Of Interest (ROIs), each measured in Hounsfield units were used as continuous-valued "objective" outcomes, and two consultant radiologists evaluated every image on a nominal scale for image quality. Because of the very skew distribution of the nominal data, these were simplified to a binary variable of "Excellent" vs every other level, results into two binary "subjective" endpoints. Examples of the ordinal scale of subjective image quality are shown in Figure \ref{fig:ordinal-scale}. Sex, age and  flow speed of injection were also collected as relevant covariates. The combination of 210 patients, 4 endpoints, 7 body locations and three energy levels resulted in a data set of size $N=17,640$ but again the Kronecker structure enabled exact inference with limited computational resources.

\section{Related Work}

Gaussian Processes have been used for decades under various names such as kriging or Bayesian kernel regression, and are related to the widely used Support Vector Machines (SVMs) \cite{hearst1998support}. Their used has increased in the past couple of decades with the advent of greater computational resources, and much research has occurred concerning their scalability to larger data sets under contemporary computational constraints. These methods often include variational inference schemes via the "inducing point" framework\cite{hensman2013gaussian}, or spectral methods to approximate the full covariance function via sampled Fourier features\cite{rahimi2007random}. The speedups possible through Kronecker structure have been used for some time, but have previously been used on structured image pixel data\cite{gilboa2014image}, drug combinations\cite{Ronneberg2021,Ronneberg2023}, spatio-temporal modelling\cite{Gelfand2004}, and multi-task regression \cite{bonilla2007multi}.

Heterogeneous multiple outputs have also received research focus \cite{moreno2018heterogeneous}: while multiple Gaussian-distributed outputs allow for analytical marginalisation of the latent variable, the presence of discrete or other non-Gaussian-distributed outputs forces the use of non-conjugate inference methods such as variational methods \cite{hoffman2013stochastic}, Expectation Propagation \cite{minka2013expectation} or sampling of the latent variable posterior. Here we opt for the latter solution, aided by the development of the Stan programming language and the underlying continuity and smoothness of all latent parameters of interest. Heterogeneity between multiple data sets has also been explored with multiple output GPs in the context of fusing randomised and observational data sets \cite{dimitriou2024data}.

\section{Results}\label{sec3}

The results describing the behaviour of the trained models are presented here. We ran experiments using 10-fold cross-validation for two GP models with and without the random effect component in the individual-level covariance function ("GP.m" and "GP.f"), and two parametric models run in brms also with or without individual-level random effects ("brms.m" and "brms.f").

\subsection{Comparison in brms}

Two brms models were used as comparison methods: ("brms.m" and "brms.f") with or without random effects at the individual level, respectively. Both models had linear fixed effects based on the covariates for each data set, i.e. for the simulated data example, each output modelled with the following model formulae:

\begin{verbatim}
 form.yi <-  yi ~ (1|p|id) + x1 + x2 + x3 # "brms.m"
 form.yi <-   yi ~ x1 + x2 + x3 # "brms.f"
\end{verbatim}

with $p$ being shared between the outputs. Either Gaussian or Binomial likelihoods were then added to the \texttt{bf} objects, and all outputs were learned jointly in a single call to \texttt{brm}. Code is available in the supplementary material.

For the first real data set, an interaction was included between time and contrast volume per unit body mass, and for the second real data set, an interaction was included between randomisation group and energy level. 

\subsection{Data Simulation}

Two distinct latent parametric functions were used to simulate the data, providing a nonlinear relationship between the outputs and the covariates:

\begin{equation}\label{eqn:simfunctions1}
    f_1 = \exp(.15x_1) - .6x_2^2 + \sin(3x_3)\\
\end{equation}
\begin{equation}\label{eqn:simfunctions2}
    f_2 = - \exp(-.15x_1) + \lvert3x_2\rvert - \cos(3x_3)
\end{equation}

Two further latent functions were defined for the binomial outputs: $f_3 = - f_1$ and $f_4 = f_2$. All four of the latent functions then had Gaussian noise of mean zero and standard deviation 0.1 added: the resulting noisy samples for the first two outputs became the observed continuous variables, while the final two noisy samples were pushed through a probit link function and rounded to generate the observed binomial variables. No random effects were included in the data simulation process. 

The covariates were sampled from $\mathcal{U}_{[-5,5]}$. 20 unique grid points were sample for $x_1$, 7 for $x_2$, and 3 for $x_3$. Combined with the four outputs and the Kronecker structure, this resulted in a total number of 1,680 unique observations.

The statistical models that included random effects at the individual level ("GP.m" and "brms.m") treated the third covariate $x_3$ as representing a measurement at the individual-level data, for the sake of comparison, but this choice is not expected to have a large influence on the results, as there was no variation in the simulated data beyond the observed covariates and shared noise.

\subsection{Losses and Testing}

The models were evaluated using 10-fold cross-validation to predict the test outputs given the training data and test covariates. Loss functions were evaluated using the mean predictive $f_{pm}$ across posterior samples and the observed data, using an quadratic/L2 loss for Gaussian outputs $y_{g}$, while for binomial outputs $y_{b}$, we used the logarithm of one minus the probability of the observed data ("the log probability of the wrong answer"), i.e. $l(y_{g}, f_{pm}) = (y_{g}- f_{pm})^2$ and $l(y_{b}, f_{pm}) = \log(\Phi(-f_{pm}*(2y_{b} - 1))$, where $\Phi$ is the Gaussian cumulative distribution function.

This resulted in distinct populations of losses per method, with one evaluated loss per output data point. These are plotted as histograms for each output in Figures \ref{Fig:SimulatedResults}, \ref{Fig:ObservationalResults},\ref{Fig:RCTResults}, with the binomial losses represented on the log scale as well as the probability scale. The differences between populations of losses for each output were tested formally using a paired Wilcoxon rank sum test, corrected for four-fold multiple testing degeneracy between methods. The p-values from the Wilcoxon tests were further supplemented by rank-biserial correlations as effect sizes. Nonparametric tests were chosen for model evaluation in order to avoid making distributional assumptions about the populations of losses, and to make the results robust to monotonic transformations in the losses.

For each of the four models on each of the data sets, we present posterior summaries of the interpretable model parameters from the first of the ten CV folds in \cref{tab:simulated.GP.f,tab:simulated.GP.m,tab:simulated.brms.f,tab:simulated.brms.m,tab:thien.GP.f,tab:thien.GP.m,tab:thien.brms.f,tab:thien.brms.m,tab:helgestad.GP.f,tab:helgestad.GP.m,tab:helgestad.brms.f,tab:helgestad.brms.m}. Considering the size of the data sets and the randomisation process used, we consider posterior summaries of a single CV fold to be representative of the population as a whole.

\subsection{Results for Simulated Data}\label{results:simulated}

\begin{figure}[b]
\centering
\begin{subfigure}{.5\textwidth}
    \centering
\includegraphics[width=.9\linewidth]{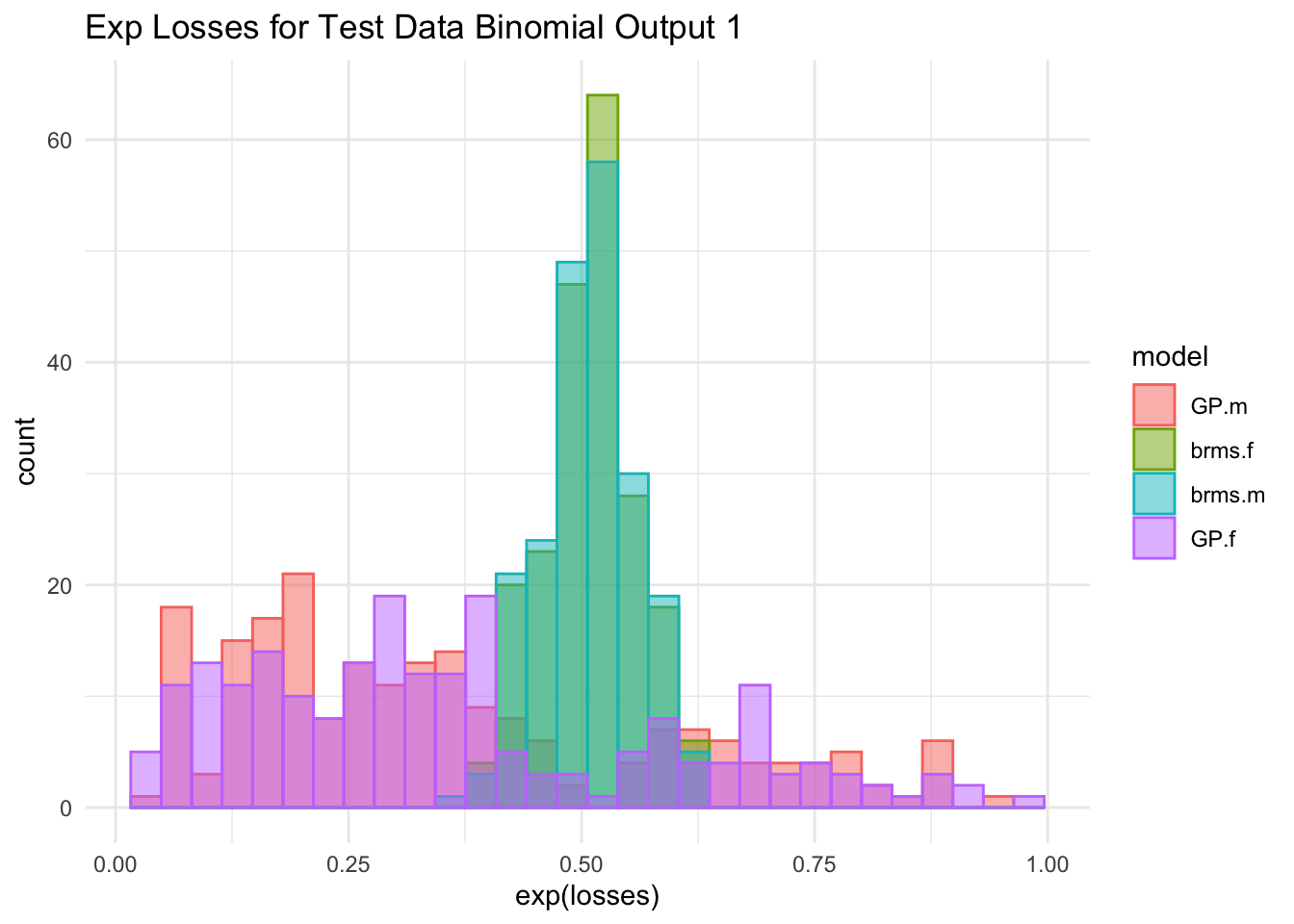} 
\end{subfigure}%
\begin{subfigure}{.5\textwidth}
    \centering
\includegraphics[width=.9\linewidth]{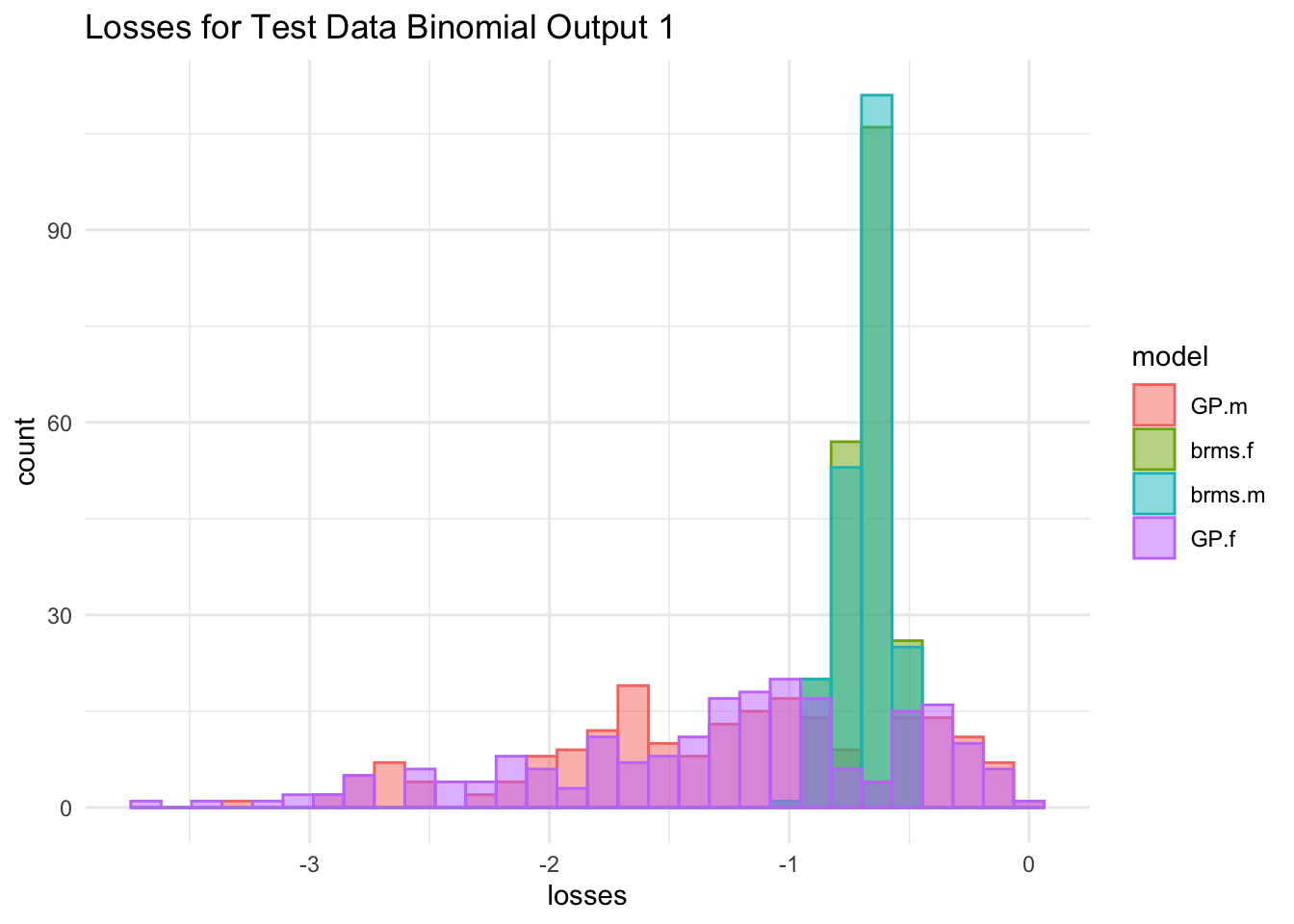}
\end{subfigure}
\begin{subfigure}{.5\textwidth}
    \centering
\includegraphics[width=.9\linewidth]{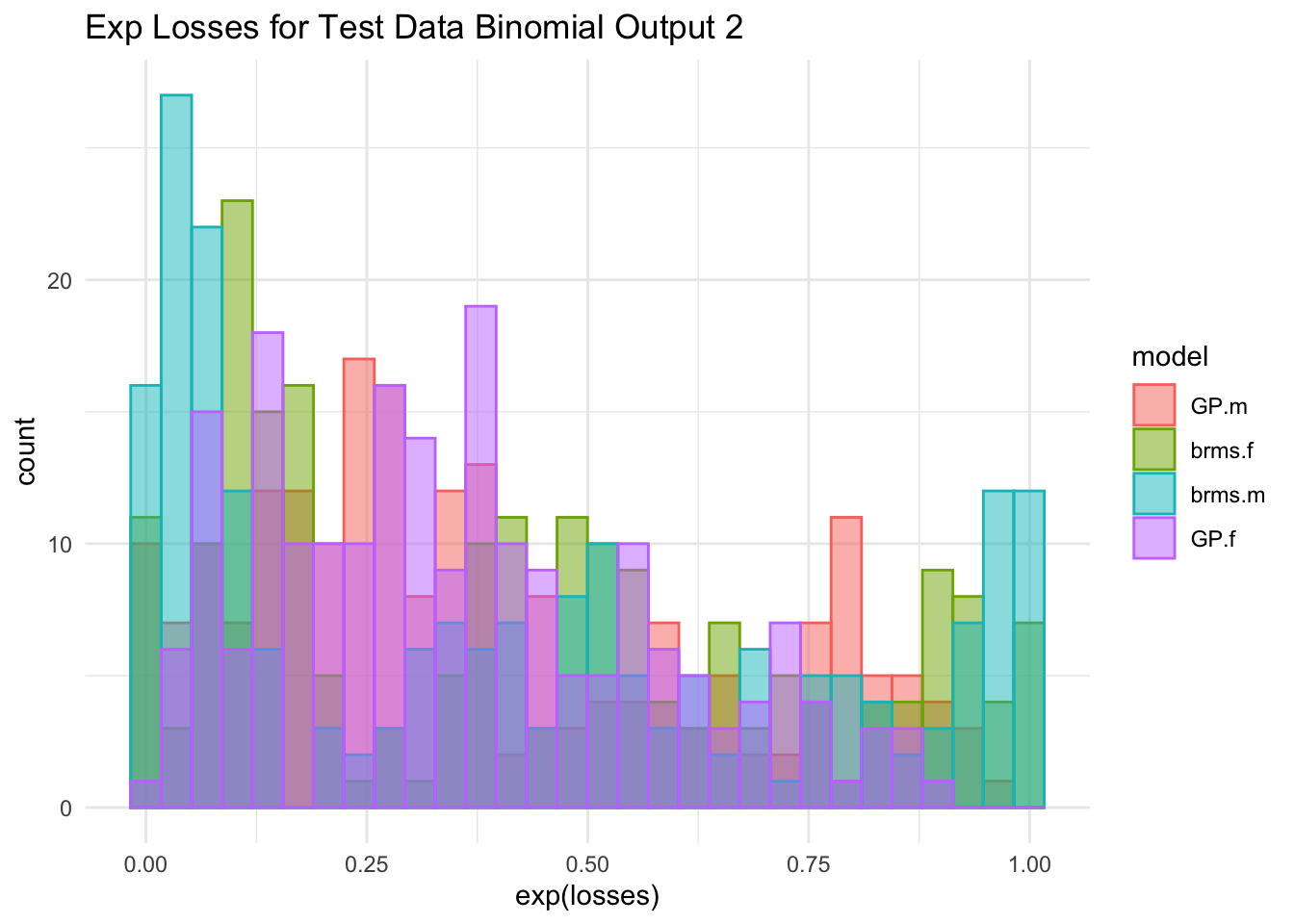} 
\end{subfigure}%
\begin{subfigure}{.5\textwidth}
    \centering
    \includegraphics[width=.9\linewidth]{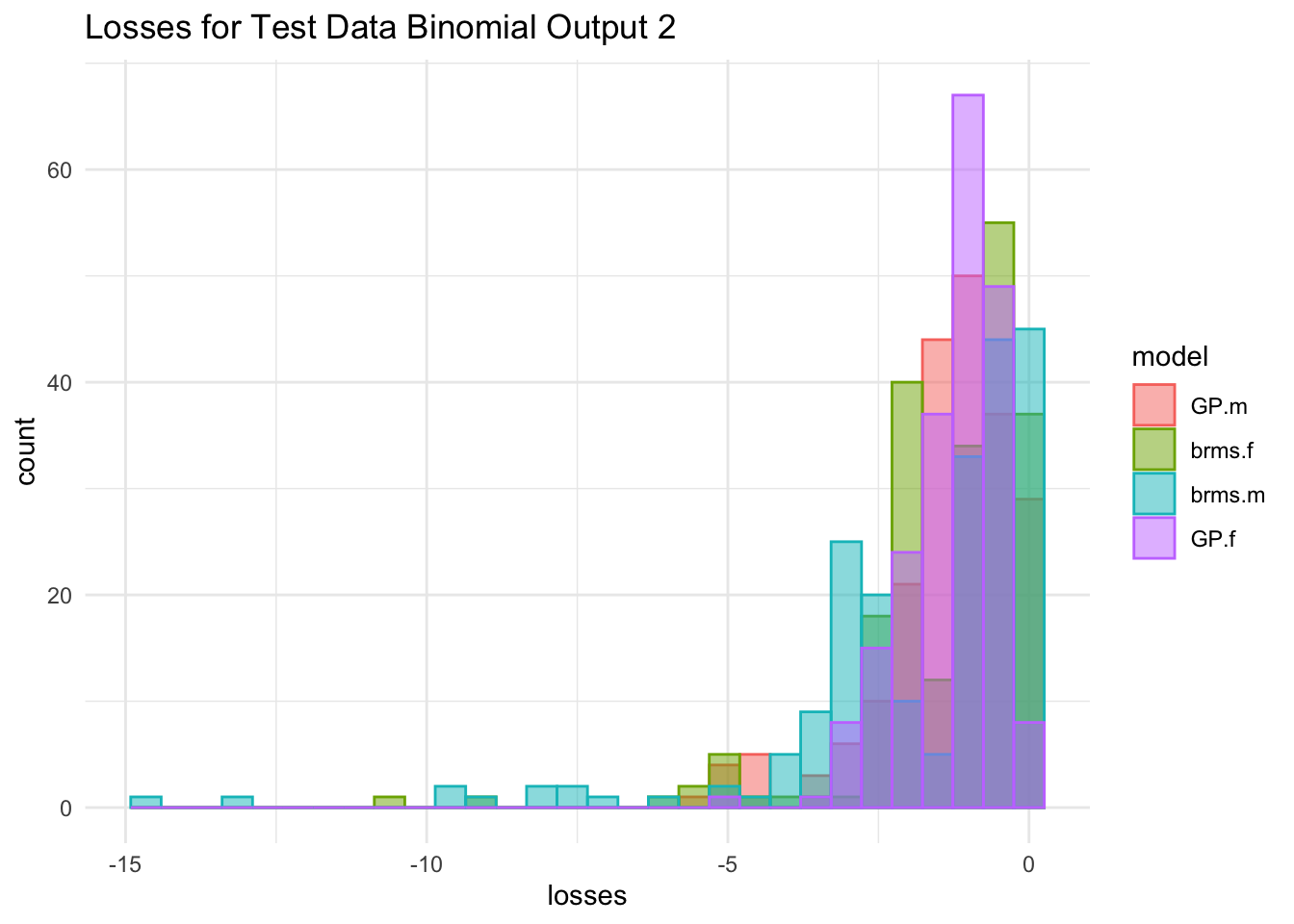}
\end{subfigure}
\begin{subfigure}{.5\textwidth}
    \centering
\includegraphics[width=.9\linewidth]{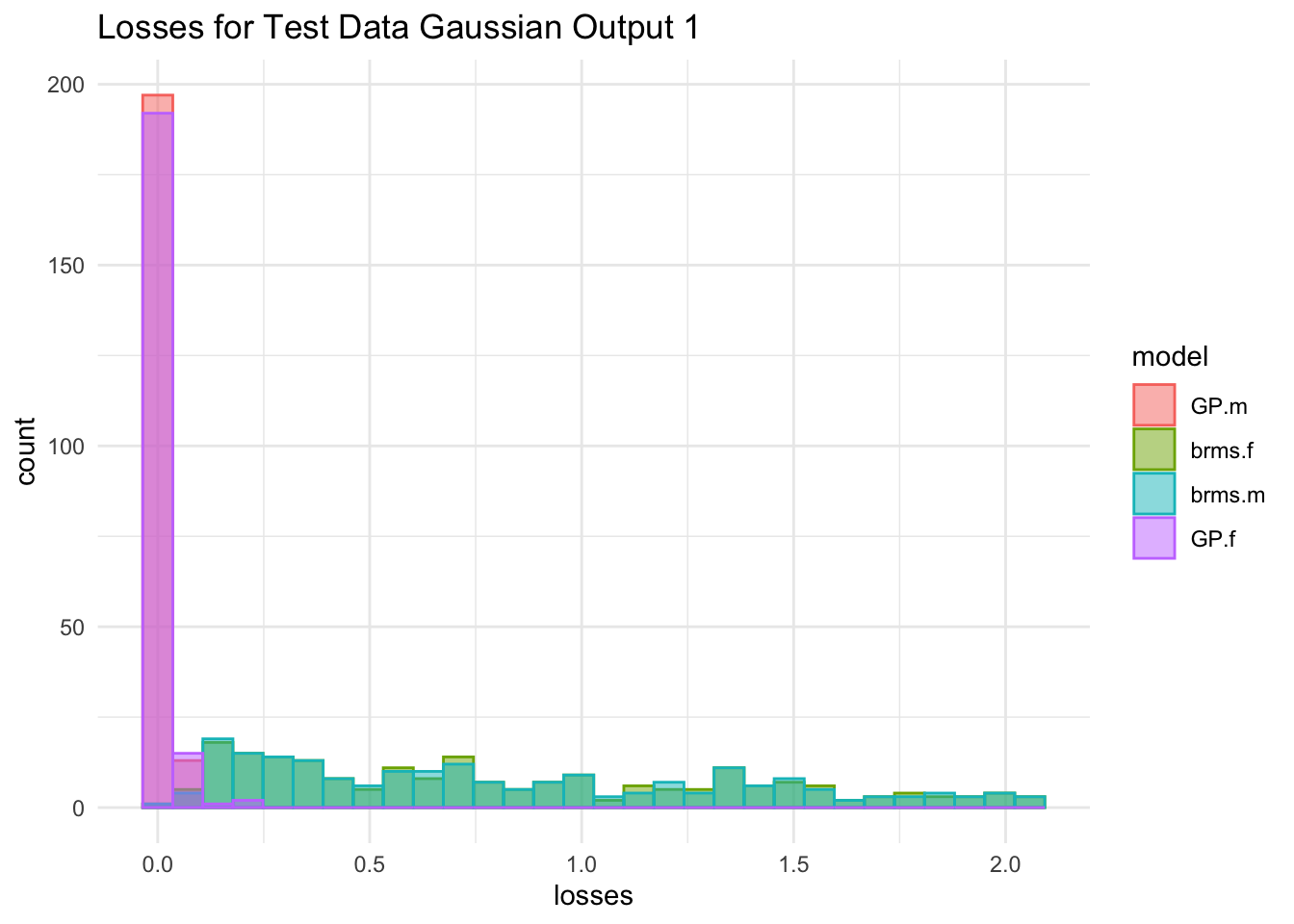} 
\end{subfigure}%
\begin{subfigure}{.5\textwidth}
    \centering
    \includegraphics[width=.9\linewidth]{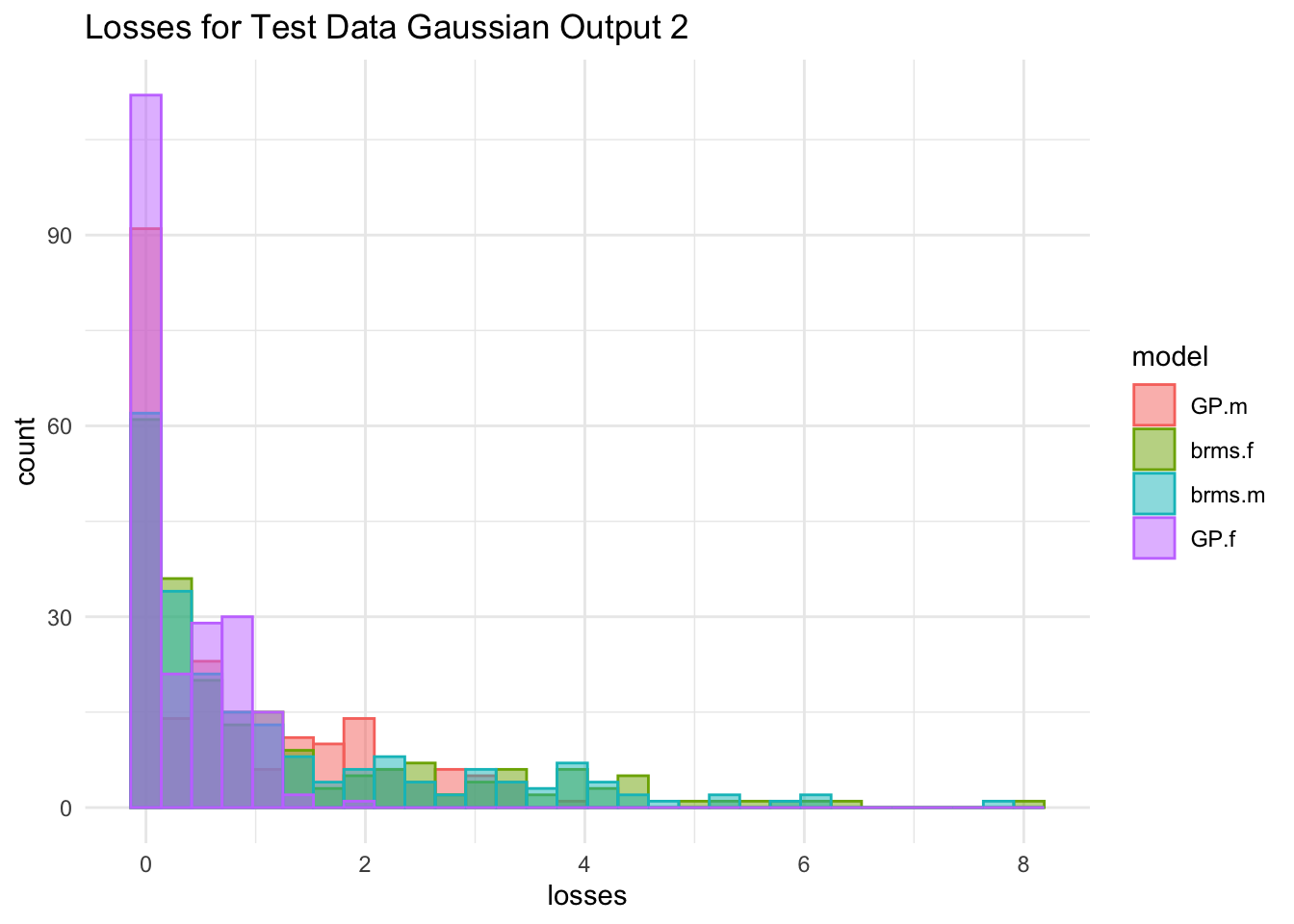}
\end{subfigure}
\caption{Figures for the simulated data example, representing histograms of losses for each of the method for each of the outputs. The uppermost two plots represent the losses on the two Gaussian-distributed outputs. Binomial outputs are plotted in the lower two rows, with those on the left on the exponentiated probability scale, and those on the right the same results plotted on the original logarithmic scale.}
\label{Fig:SimulatedResults}
\end{figure}

We see the predictive loss results in Table \ref{Tab:SimulatedResults} and Figure \ref{Fig:SimulatedResults}. We see that the brms model effectively fails and returns samples from the prior on binomial output 1 and Gaussian output 1, which are derived from the same latent parametric function. The GP models consequently registered significantly lower losses with large effect sizes. We see a small and possibly spurious effect of the GP models appearing to not predict around the prior at $p=0.5$. For the second continuous output, we again see the GP models outperforming the brms models, with the GP.f performing substantially better than GP.m. For the second binomial output, we see that the brms models get many of the labels correct with high confidence, but also many of the labels wrong with high confidence. Consequently, the results here are more mixed, with the only significant differences being brms.m outperforming both GP.m and brms.f. We would expect from the absolute value present in \ref{eqn:simfunctions2} that the smooth GP model would find the second and fourth endpoints more challenging to model accurately, relative to the first and third.

Parameter posterior summaries are presented for all four models trained on this data set in \cref{tab:simulated.GP.f,tab:simulated.GP.m,tab:simulated.brms.f,tab:simulated.brms.m}.

\begin{table}
\centering
\begin{tabular}{cc|cccc|cccc}
& & \multicolumn{4}{c}{GAUSSIAN} & \multicolumn{4}{c}{BINOMIAL} \\
& & GP.m & brms.f & brms.m & GP.f & GP.m & brms.f & brms.m & GP.f \\ \hline
\multirow{4}{*}{ 1 } & GP.m & - & -1.000 & -1.000 & -0.091 & - & -0.778 & -0.778 & 0.077 \\
& brms.f & $<.001$ & - & 0.044 & 1.000 & $<.001$ & - & -0.015 & 0.769 \\
& brms.m & $<.001$ & 0.582 & - & 1.000 & $<.001$ & 0.852 & - & 0.769 \\
& GP.f & 0.508 & $<.001$ & $<.001$ & - & 0.664 & $<.001$ & $<.001$ & - \\ \hline
\multirow{4}{*}{ 2 } & GP.m & - & -0.608 & -0.602 & 0.534 & - & -0.083 & 0.255 & -0.074 \\
& brms.f & $<.001$ & - & 0.182 & 0.777 & 0.892 & - & 0.606 & 0.018 \\
& brms.m & $<.001$ & 0.022 & - & 0.781 & 0.007 & $<.001$ & - & -0.197 \\
& GP.f & $<.001$ & $<.001$ & $<.001$ & - & 0.892 & 0.892 & 0.053 & - \\
\end{tabular}
\caption{Results for the simulated data example, comparing between populations of losses for each modelling method, with the first Gaussian output in the upper left, the first Binomial to the upper right, the second Gaussian to the lower left, and the second Binomial to the lower right. Within each square, the left-lower-triangular results are the p-values from each wilcoxon rank sum test, while the right-upper-triangular results are the rank-biserial correlations representing effect sizes.}
\label{Tab:SimulatedResults}
\end{table}

\subsection{Observational Time-Series Data}\label{results:thien}

\begin{figure}[b]
\centering
\begin{subfigure}{.5\textwidth}
    \centering
\includegraphics[width=.9\linewidth]{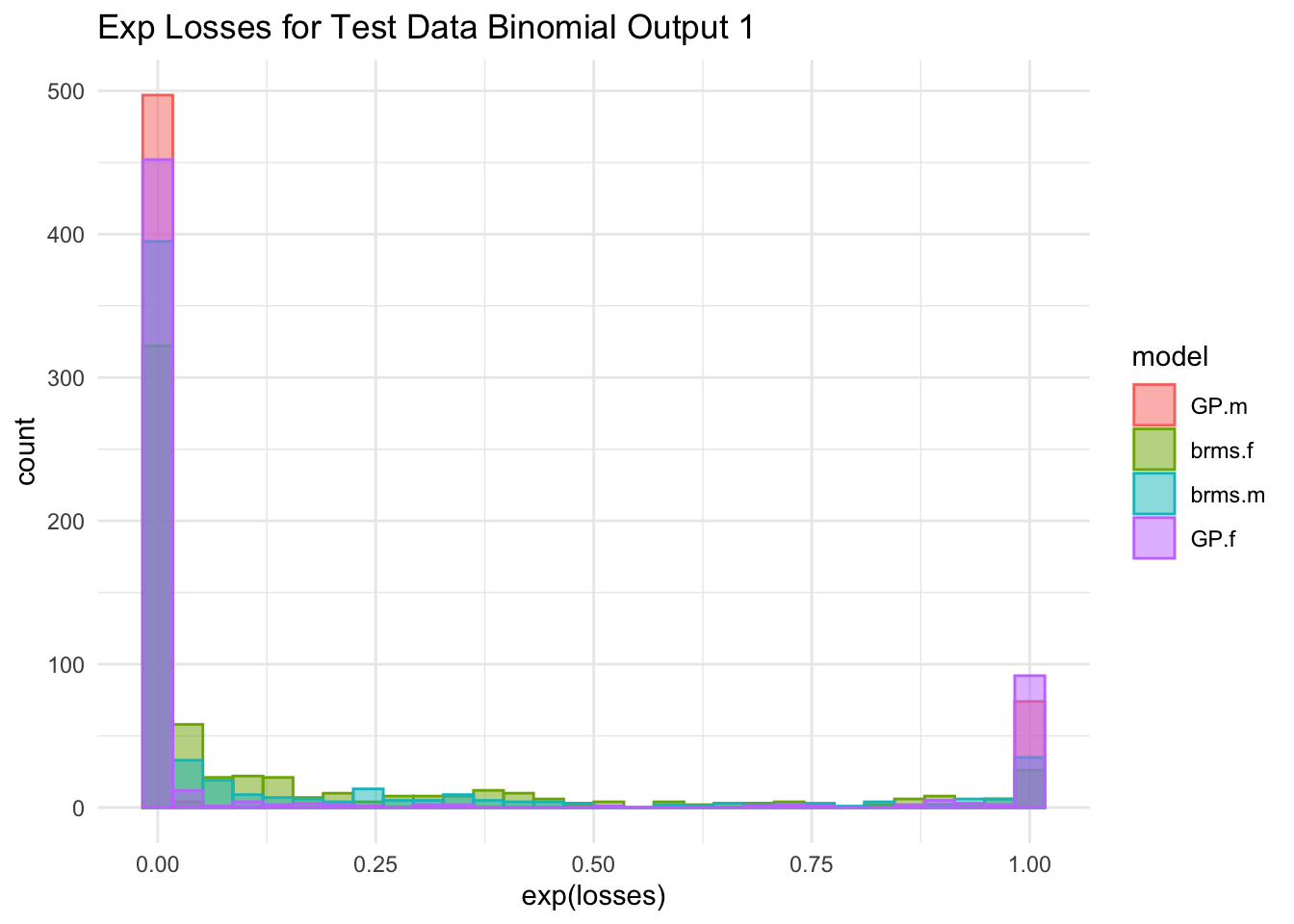} 
\end{subfigure}%
\begin{subfigure}{.5\textwidth}
    \centering
\includegraphics[width=.9\linewidth]{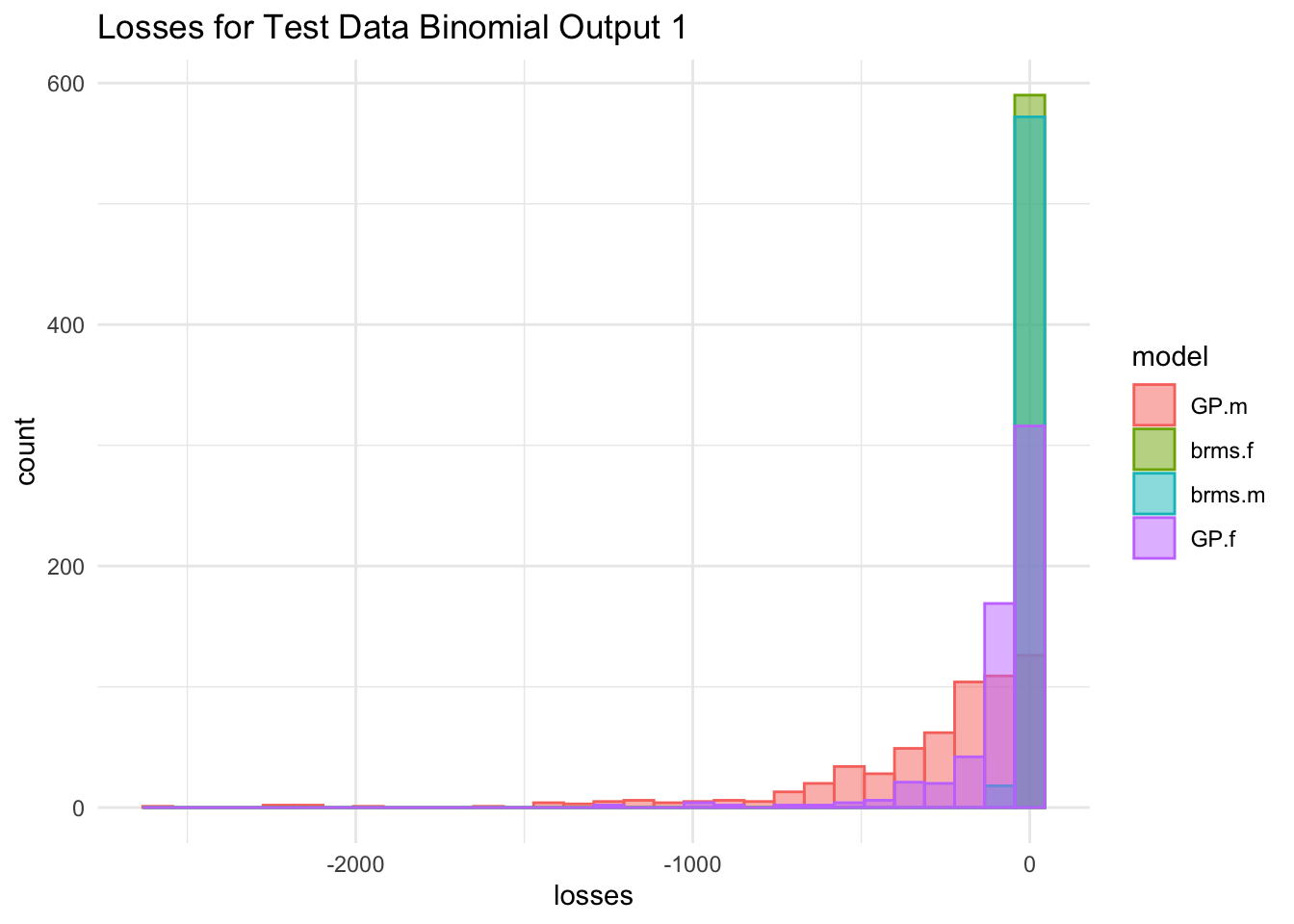}
\end{subfigure}
\begin{subfigure}{.5\textwidth}
    \centering
\includegraphics[width=.9\linewidth]{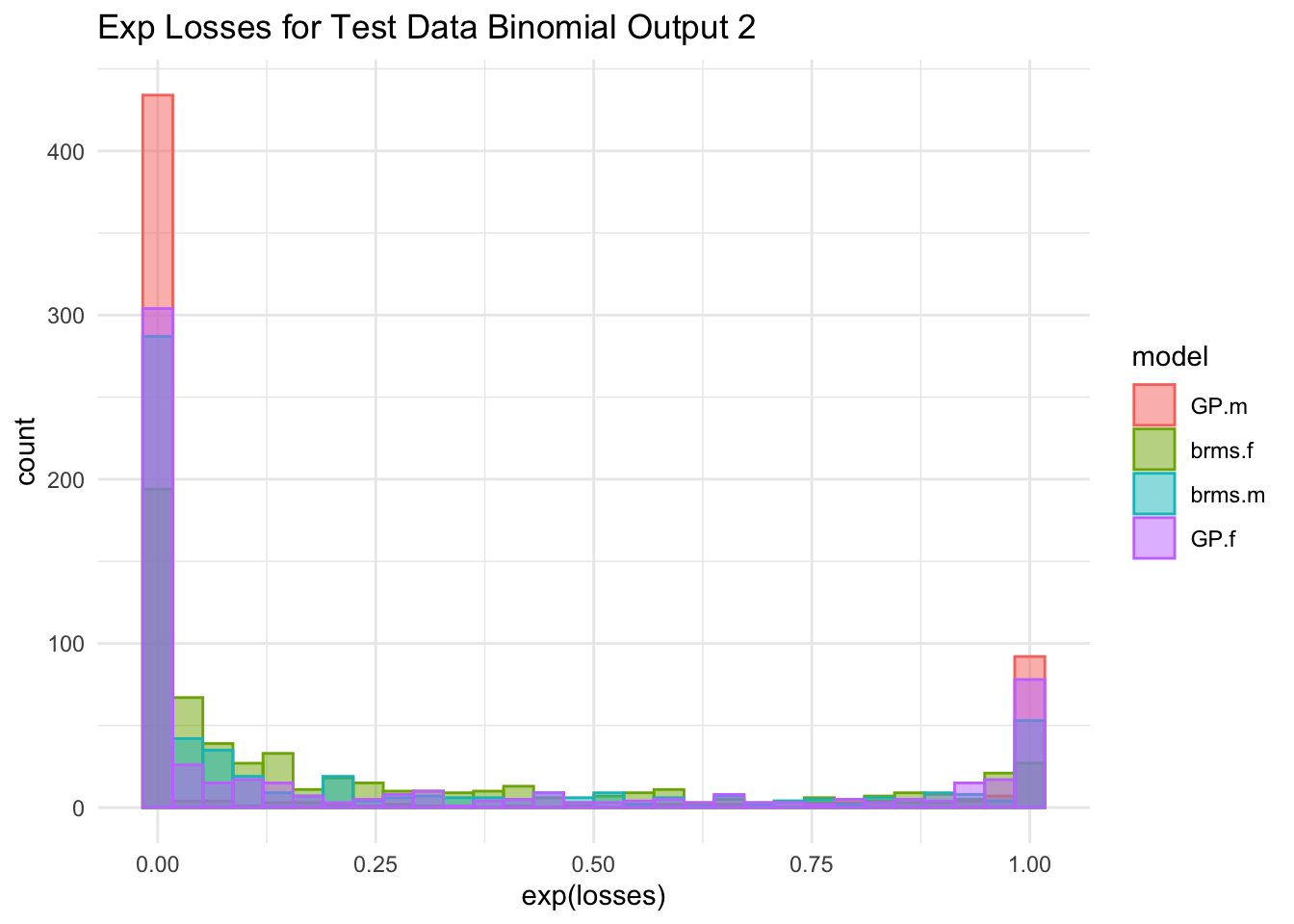} 
\end{subfigure}%
\begin{subfigure}{.5\textwidth}
    \centering
    \includegraphics[width=.9\linewidth]{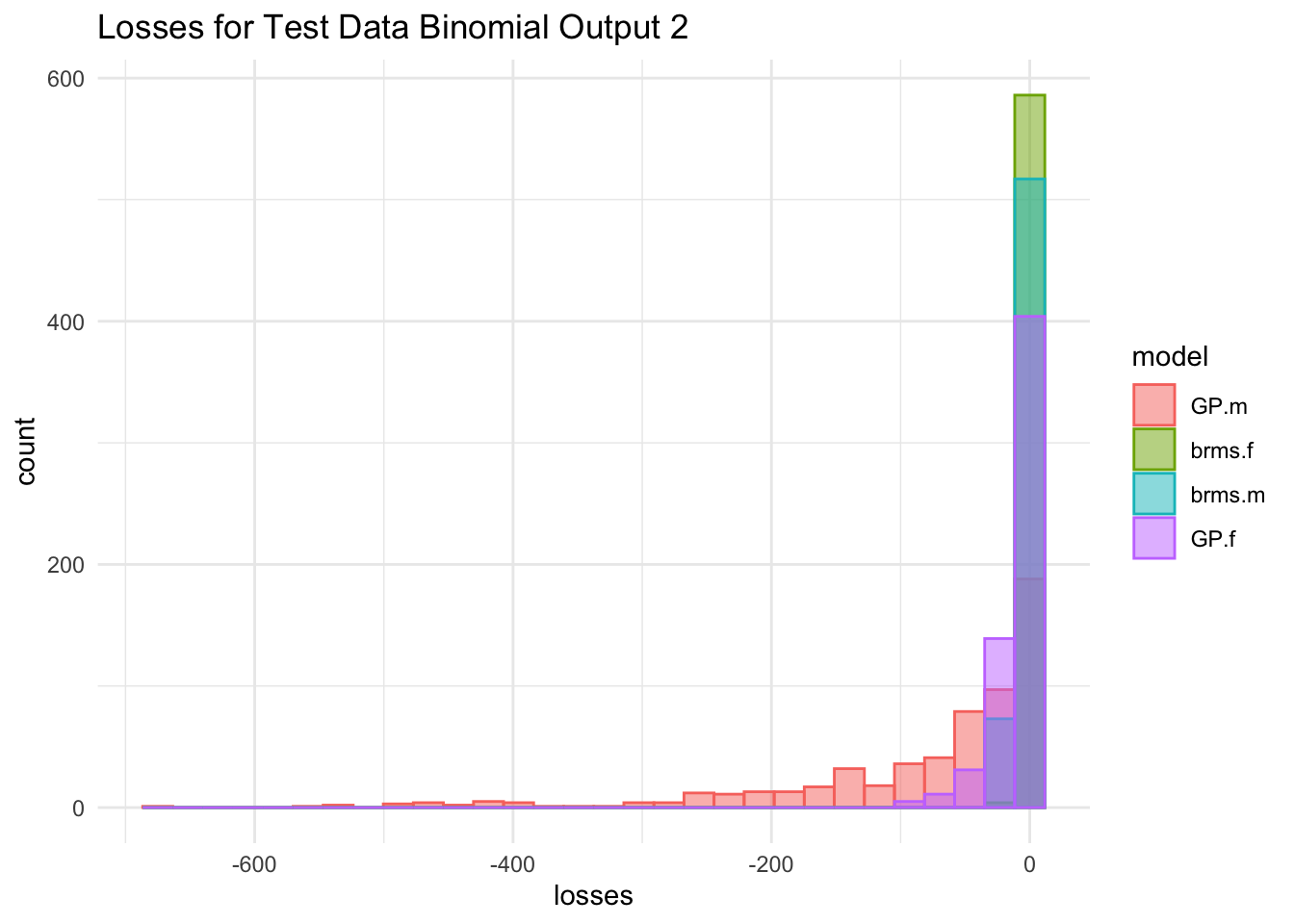}
\end{subfigure}
\begin{subfigure}{.5\textwidth}
    \centering
\includegraphics[width=.9\linewidth]{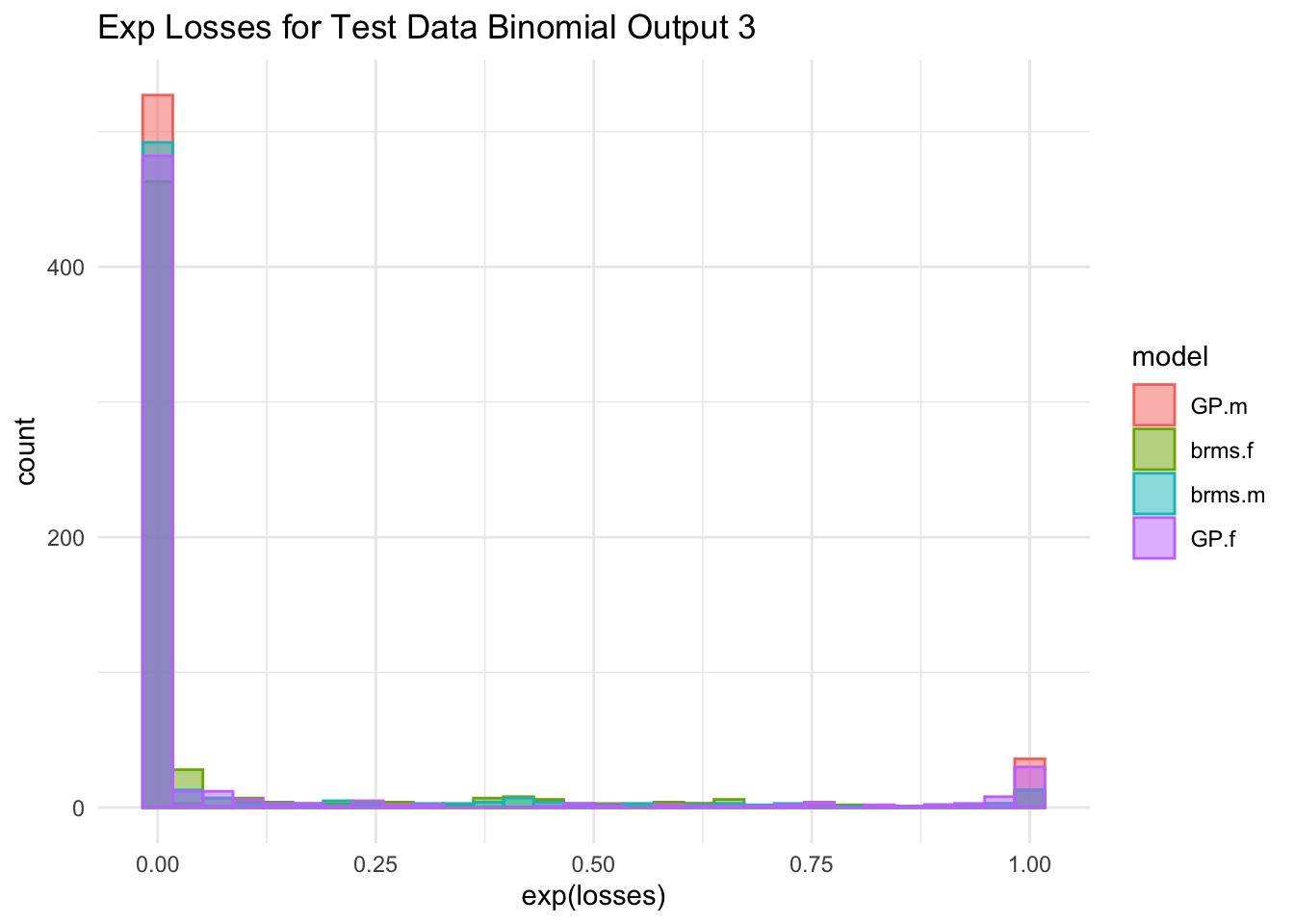} 
\end{subfigure}%
\begin{subfigure}{.5\textwidth}
    \centering
    \includegraphics[width=.9\linewidth]{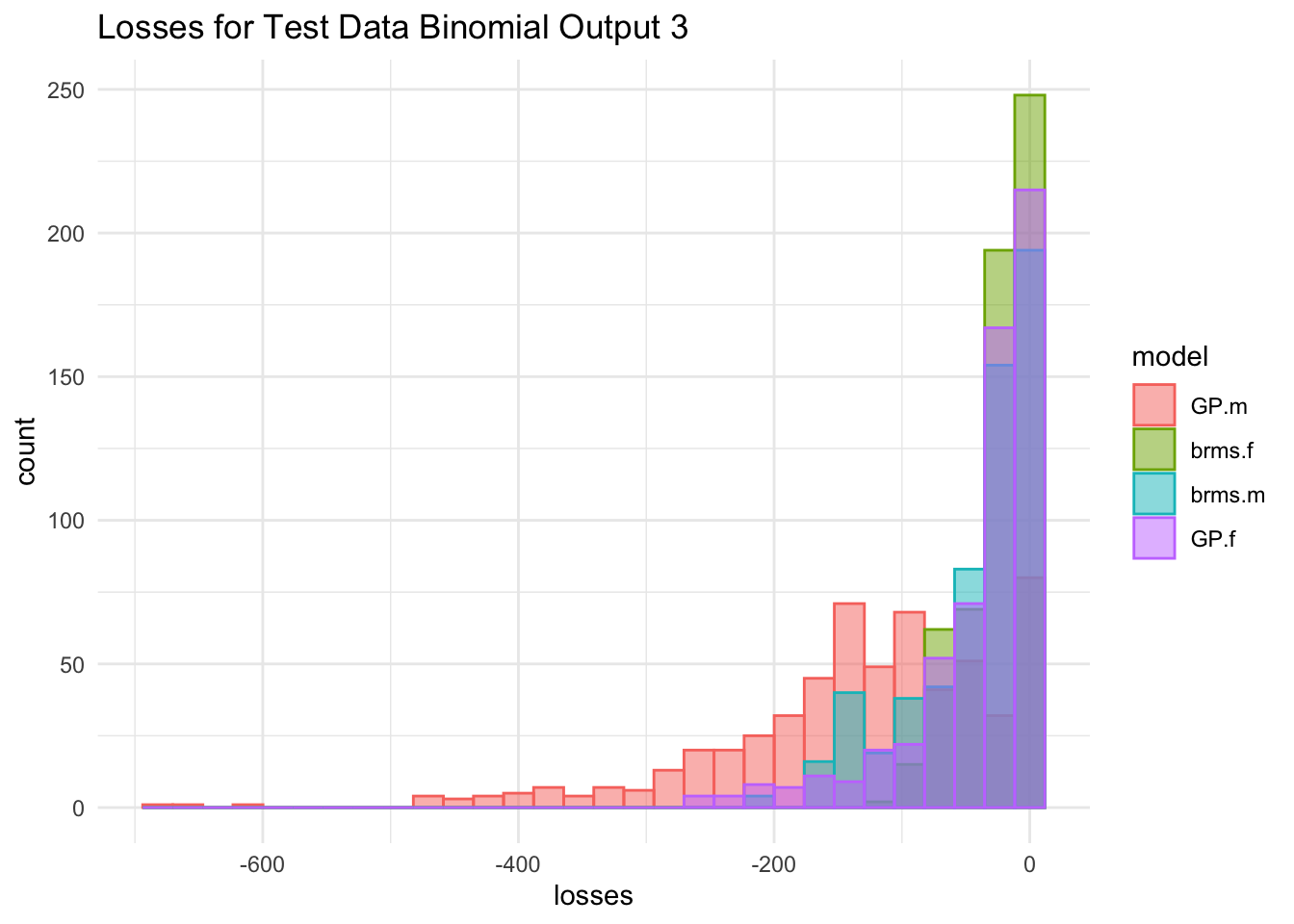}
\end{subfigure}
\begin{subfigure}{.5\textwidth}
    \centering
\includegraphics[width=.9\linewidth]{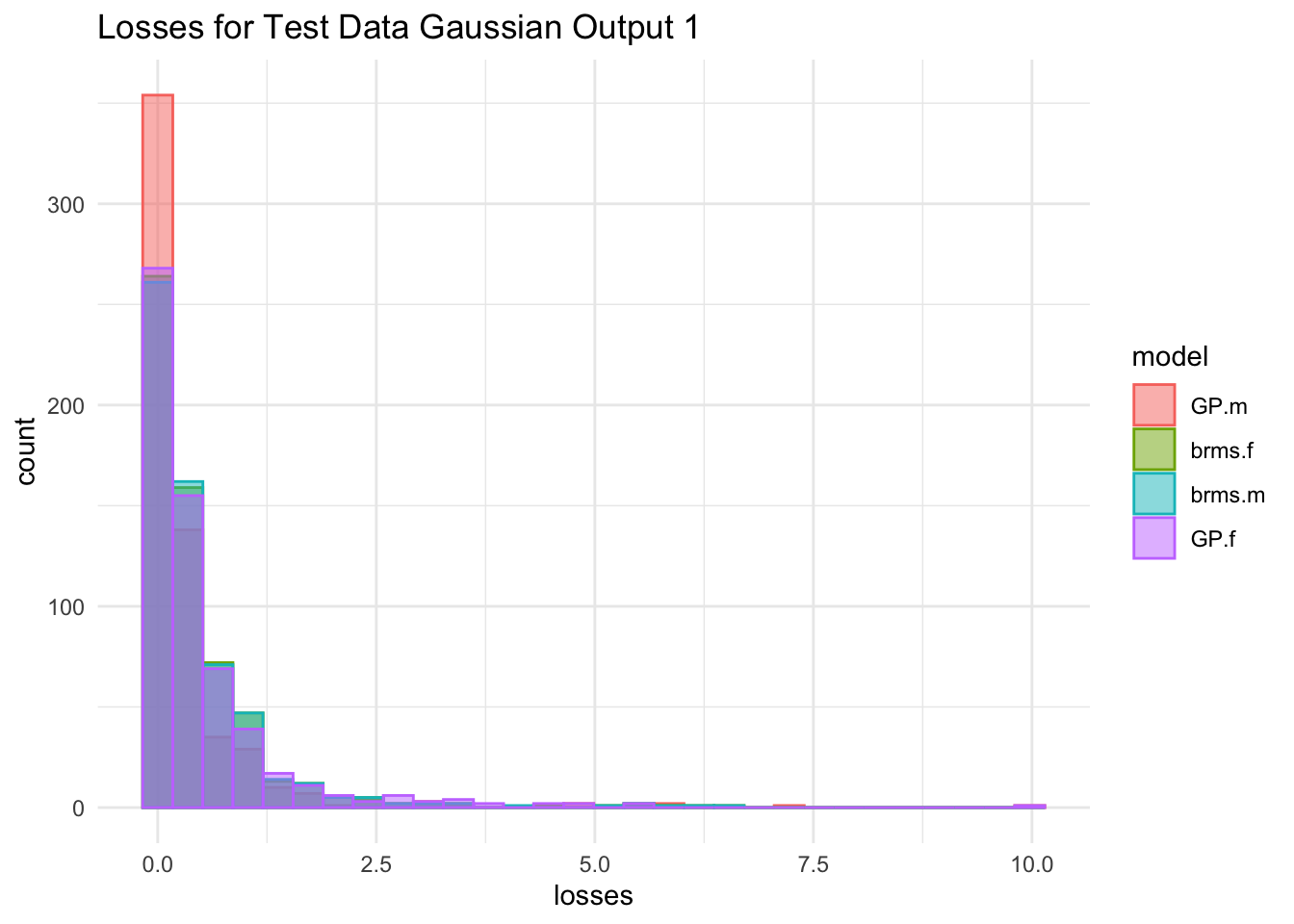}
\end{subfigure} \caption[h]{Figures for the observational time series data set, representing histograms of losses for each of the method for each of the outputs. Binomial outputs are plotted in the first three rows, with those on the left on the exponentiated probability scale, and those on the right the same results plotted on the original logarithmic scale. The lowest-left plot represents the losses on the one continuous output }
\label{Fig:ObservationalResults}
\end{figure}

We see the predictive loss results for the first real data set in Table \ref{Tab:ObservationalResults} and Figure \ref{Fig:ObservationalResults}. We see that all of the methods were confidently correct in their predictions of labels for the binomial outputs, suggesting that this is a relatively easy classification problem. That said, when considering the ranking of the predictions, we see that GP.m model outperforms the other methods on the binomial outputs, the GP.f model mostly outperforms the brms models, and the brm.m model outperforms the brms.f model. We see a small number of extremely confident predictions from the GP.m model, especially for output 1, which is a concern even if they are correct, as the probabilities are unlikely to be well-calibrated.

When considering the continuous output, we see more mixed results between the methods, with the only clearly significant result being that the GP.m model performs uniformly better than the other methods, whose losses are otherwise indistinguishable.

Parameter posterior summaries are presented for all four models trained on this data set in \cref{tab:thien.GP.f,tab:thien.GP.m,tab:thien.brms.f,tab:thien.brms.m}.

\begin{table}
\centering
\begin{tabular}{cc|cccc|cccc|cc}
& & GP.m & brms.f & brms.m & GP.f & GP.m & brms.f & brms.m & GP.f & & \\ \hline
\multirow{4}{*}{ G.1 } & GP.m & - & -0.390 & -0.389 & -0.343 & - & -0.878 & -0.859 & -0.839 & GP.m & \multirow{4}{*}{ B.2 } \\
& brms.f & $<.001$ & - & 0.014 & -0.005 & $<.001$ & - & 0.885 & 0.552 & brms.f & \\
& brms.m & $<.001$ & 1.000 & - & -0.006 & $<.001$ & $<.001$ & - & 0.327 & brms.m & \\
& GP.f & $<.001$ & 1.000 & 1.000 & - & $<.001$ & $<.001$ & $<.001$ & - & GP.f & \\ \hline
\multirow{4}{*}{ B.1 } & GP.m & - & -0.954 & -0.953 & -0.750 & - & -0.965 & -0.837 & -0.879 & GP.m & \multirow{4}{*}{ B.3 } \\
& brms.f & $<.001$ & - & 0.962 & 0.833 & $<.001$ & - & 0.987 & 0.312 & brms.f & \\
& brms.m & $<.001$ & $<.001$ & - & 0.737 & $<.001$ & $<.001$ & - & -0.106 & brms.m & \\
& GP.f & $<.001$ & $<.001$ & $<.001$ & - & $<.001$ & $<.001$ & $<.001$ & - & GP.f & \\
\end{tabular}
\caption{Results for the observational time series data set, comparing between populations of losses for each modelling method, with the Gaussian output in the upper-left, and the three Binomial outputs in the lower-left, upper-right, and lower-right. Within each square, the left-lower-triangular results are the p-values from each wilcoxon rank sum test, while the right-upper-triangular results are the rank-biserial correlations representing effect sizes.}
\label{Tab:ObservationalResults}
\end{table}

\subsection{Randomised Control Trial Data}\label{results:helgestad}

\begin{figure}[b]
\centering
\begin{subfigure}{.5\textwidth}
    \centering
\includegraphics[width=.9\linewidth]{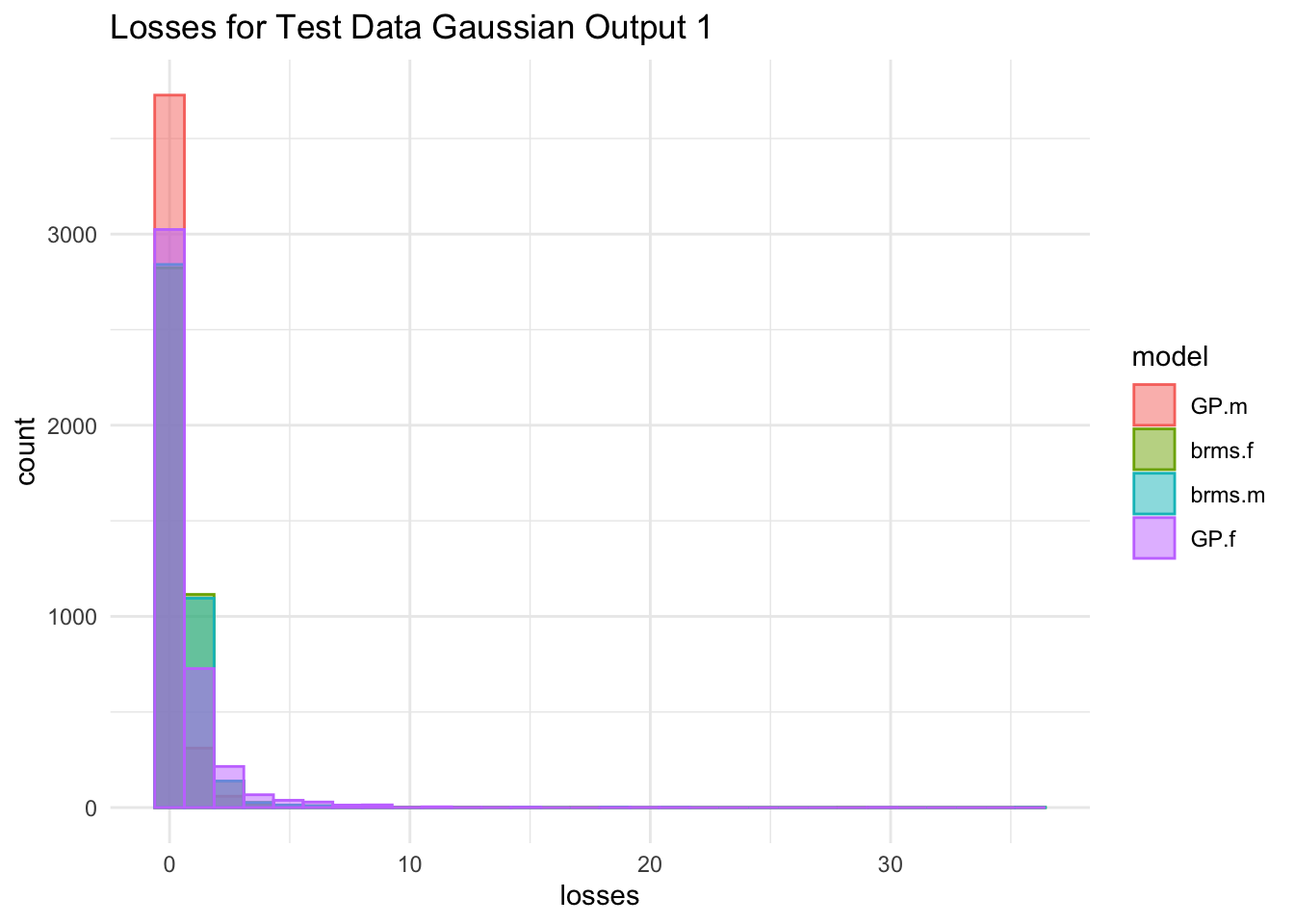} 
\end{subfigure}%
\begin{subfigure}{.5\textwidth}
    \centering
\includegraphics[width=.9\linewidth]{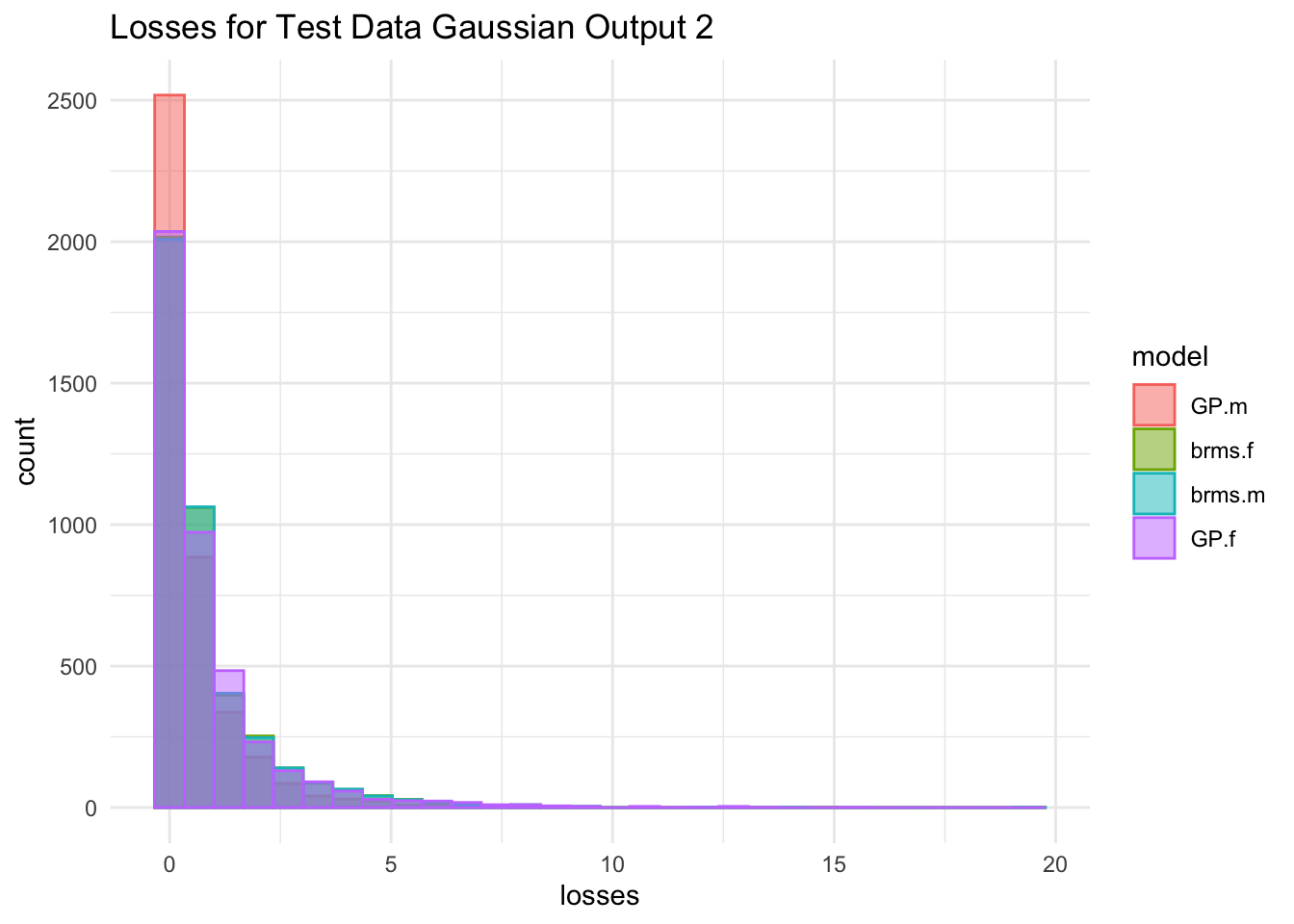}
\end{subfigure}
\begin{subfigure}{.5\textwidth}
    \centering
\includegraphics[width=.9\linewidth]{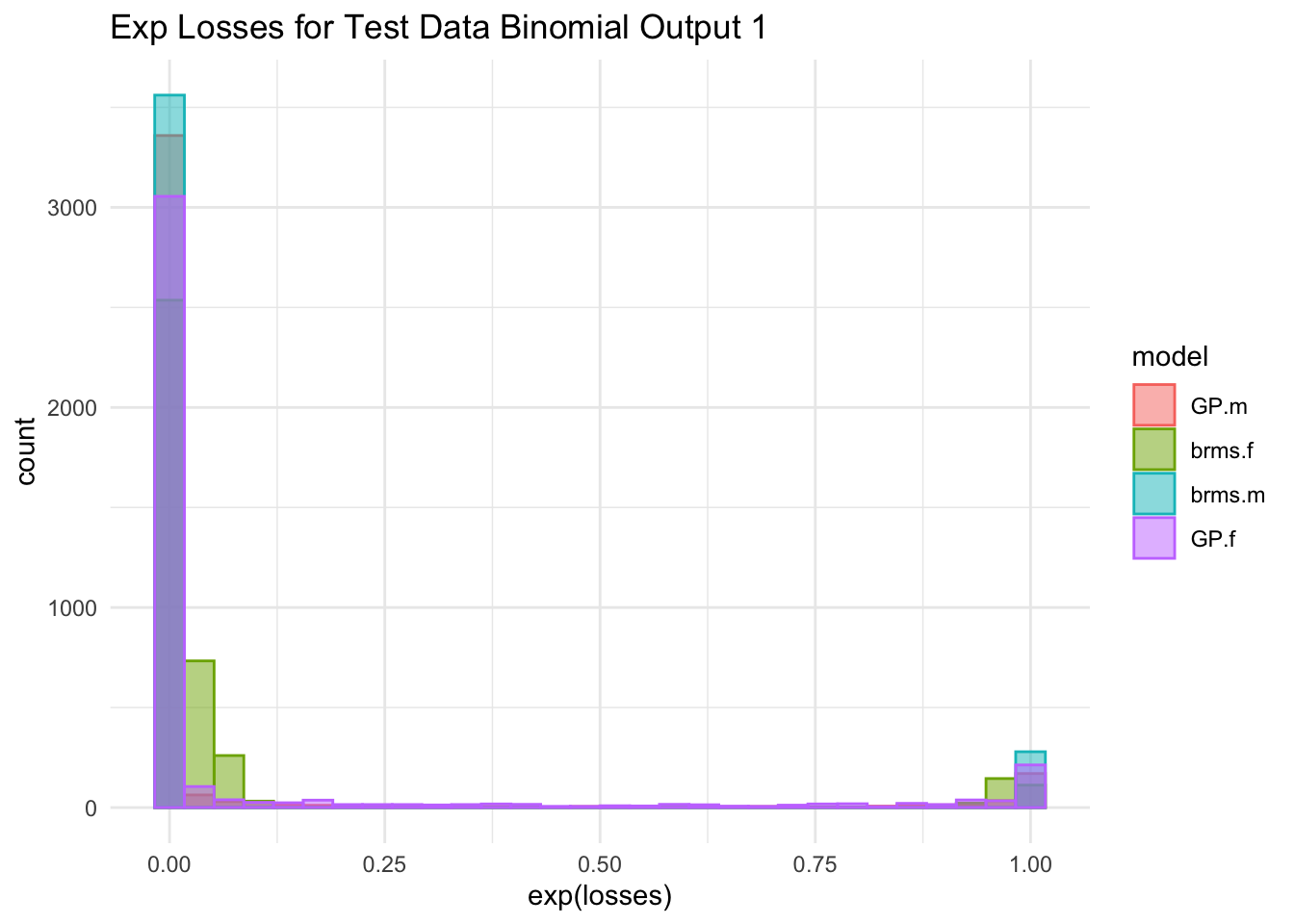} 
\end{subfigure}%
\begin{subfigure}{.5\textwidth}
    \centering
    \includegraphics[width=.9\linewidth]{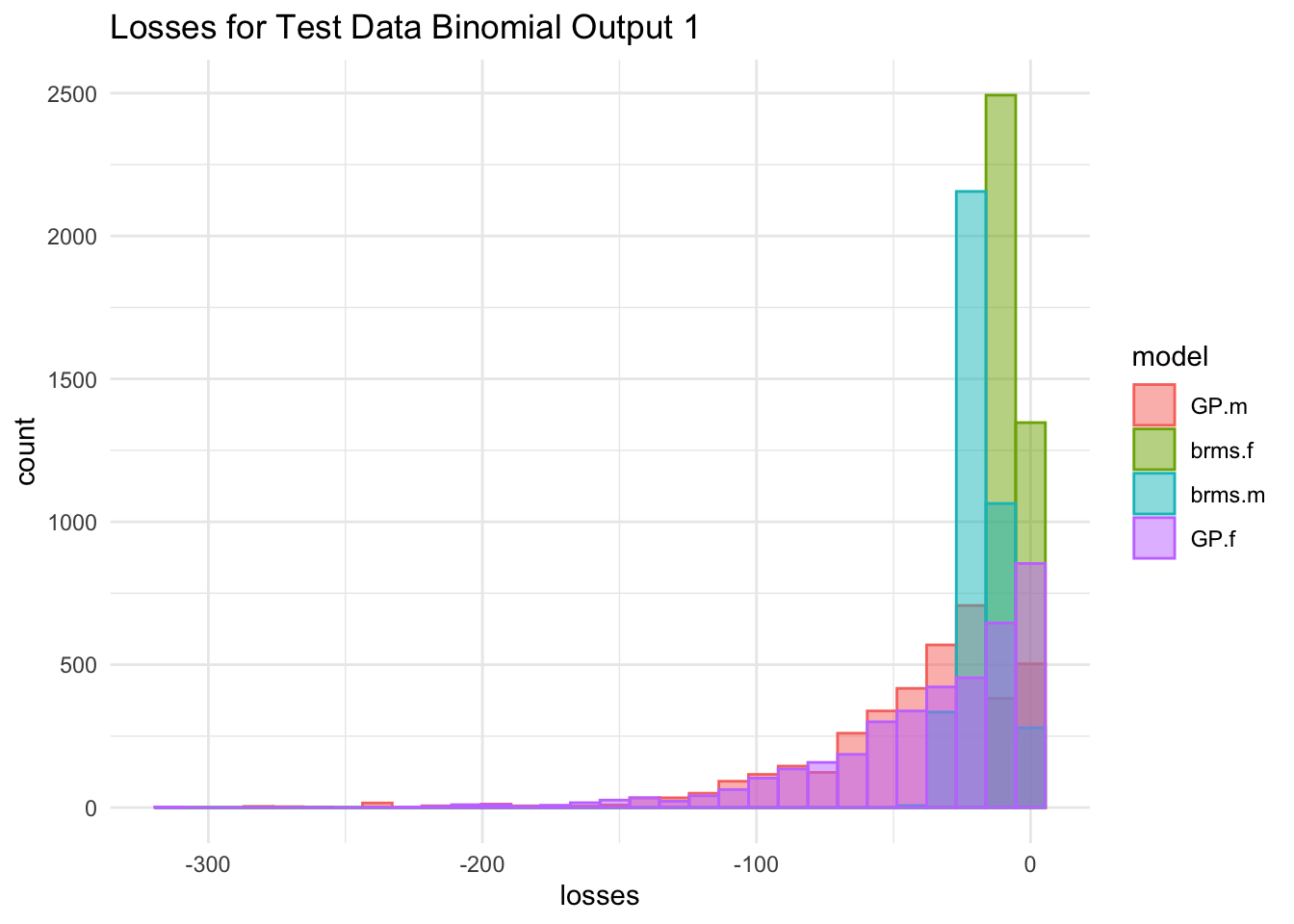}
\end{subfigure}
\begin{subfigure}{.5\textwidth}
    \centering
\includegraphics[width=.9\linewidth]{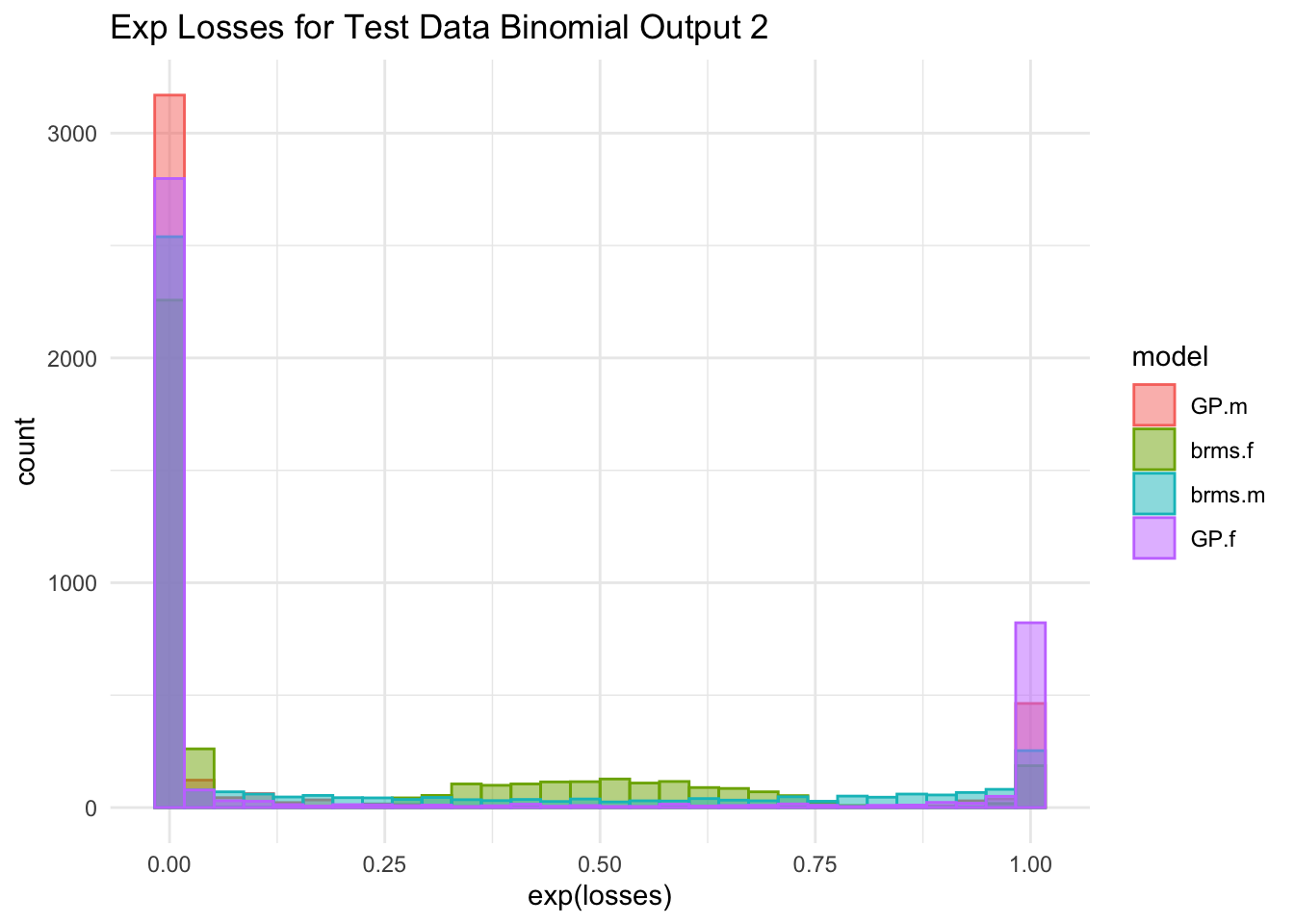} 
\end{subfigure}%
\begin{subfigure}{.5\textwidth}
    \centering
    \includegraphics[width=.9\linewidth]{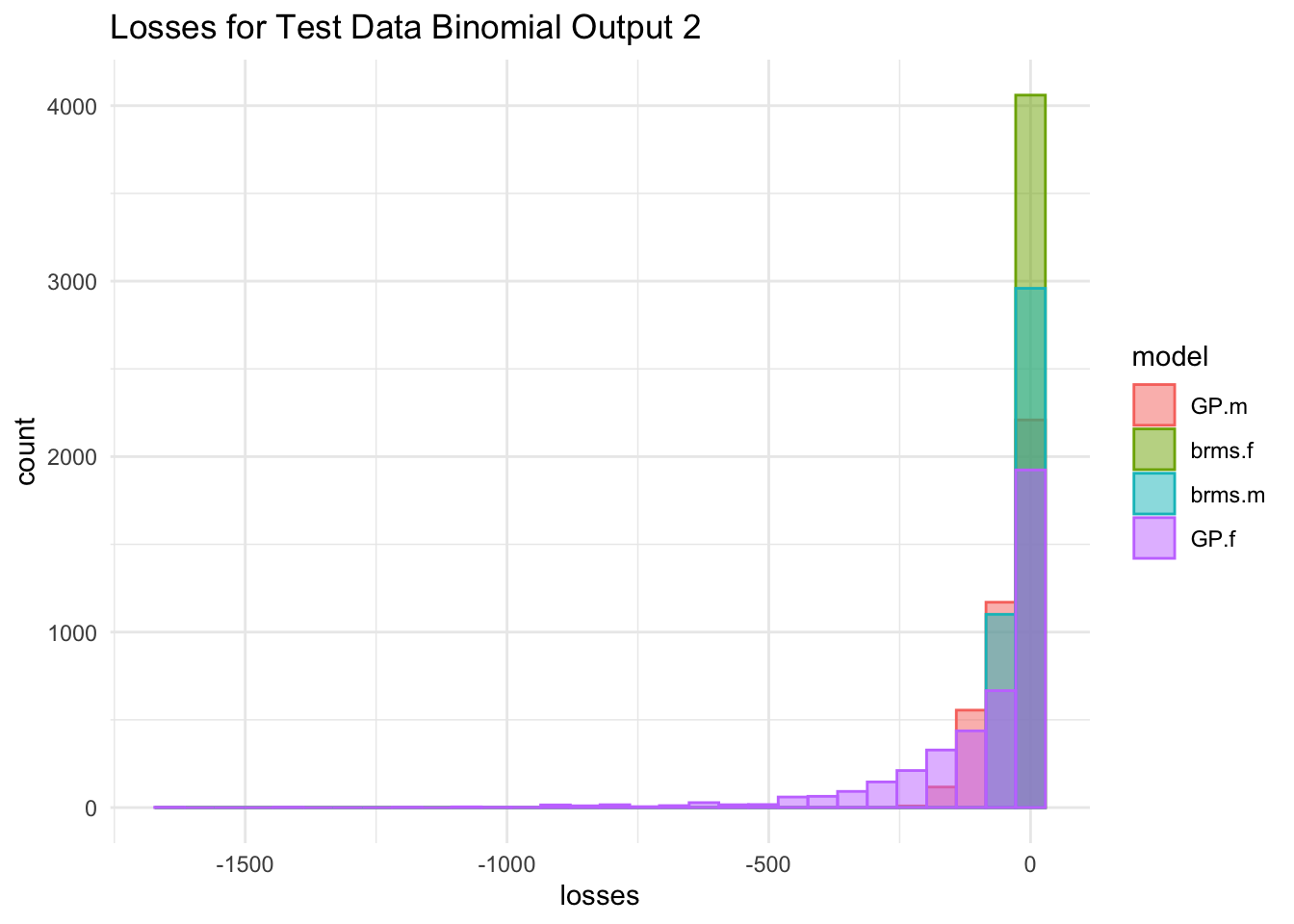}
\end{subfigure}
\caption{Figures for the RCT data example, representing histograms of losses for each of the method for each of the outputs. Binomial outputs are plotted in the first two rows, with those on the left on the exponentiated probability scale, and those on the right the same results plotted on the original logarithmic scale. The lowest two plots represent the losses on the two Gaussian-distributed outputs.}
\label{Fig:RCTResults}
\end{figure}

We see the predictive loss results for the second real data set in Table \ref{Tab:RCTResults} and Figure \ref{Fig:RCTResults}. We again see that all four of the methods predict the correct label confidently most of the time, again suggesting that this is a relatively easy prediction problem. The brms.f method appears to return a small number of predictions from the prior for the second binomial output. We see from the hypothesis testing that the GP.m model outperforms the other three methods for the first binomial output (with a small effect size relative to GP.f), whereas the GP.f model outperforms the other methods for the second binomial output.

For the continuous outputs, the GP.m model outperforms the three other methods in prediction, while the GP.f model performs slightly better or equally well as the brms models.

Parameter posterior summaries are presented for all four models trained on this data set in \cref{tab:helgestad.GP.f,tab:helgestad.GP.m,tab:helgestad.brms.f,tab:helgestad.brms.m}.

\begin{table}
\centering
\begin{tabular}{cc|cccc|cccc}
& & \multicolumn{4}{c}{GAUSSIAN} & \multicolumn{4}{c}{BINOMIAL} \\
& & GP.m & brms.f & brms.m & GP.f & GP.m & brms.f & brms.m & GP.f \\ \hline
\multirow{4}{*}{ 1 } & GP.m & - & -0.593 & -0.593 & -0.444 & - & -0.974 & -0.774 & -0.197 \\
& brms.f & $<.001$ & - & -0.017 & 0.107 & $<.001$ & - & 0.989 & 0.859 \\
& brms.m & $<.001$ & 0.352 & - & 0.108 & $<.001$ & $<.001$ & - & 0.501 \\
& GP.f & $<.001$ & $<.001$ & $<.001$ & - & $<.001$ & $<.001$ & $<.001$ & - \\ \hline
\multirow{4}{*}{ 2 } & GP.m & - & -0.357 & -0.359 & -0.360 & - & -0.940 & -0.681 & 0.417 \\
& brms.f & $<.001$ & - & -0.058 & -0.007 & $<.001$ & - & 0.878 & 0.817 \\
& brms.m & $<.001$ & 0.003 & - & -0.004 & $<.001$ & $<.001$ & - & 0.644 \\
& GP.f & $<.001$ & 1.000 & 1.000 & - & $<.001$ & $<.001$ & $<.001$ & - \\
\end{tabular}
\caption{Results for the RCT data set, comparing between populations of losses for each modelling method, with the Gaussian output in the upper-left, and the three Binomial outputs in the lower-left, upper-right, and lower-right. Within each square, the left-lower-triangular results are the p-values from each wilcoxon rank sum test, while the right-upper-triangular results are the rank-biserial correlations representing effect sizes.}
\label{Tab:RCTResults}
\end{table}

\section{Discussion}

In this work, we have demonstrated that the repeated measurements used in applied medical research can be used to scale complex Bayesian nonparametric Gaussian Process models to practical clinical questions. The more complex nonparametric models have showed to have increased predictive ability compared to the widely-used parametric models, with one of the two GP models considered demonstrating the best predictive ability in every output of the real data sets considered. Consequently, pursuing the GP modelling framework is advantageous, if achieving the best possible model specification is a concern.

Model specification is particularly relevant in medical work when there are specific clinical questions or causal hypotheses to investigate: a poorly-specified model is prone to bias estimates of the parameters of interest and hence possibly give misleading results. The question remains as to how to represent clinical hypotheses within a GP framework: while in principle parametric regression models with finite linear combinations of features have corresponding covariance function representations, it may not be straightforward to interpret covariance function variances and lengthscales in a practical way. Preexisting familiarity with ANOVA and similar models that explain contributions to the variance rather than the functional form of the mean may be a helpful reference point.

For randomised data, in which one of the covariate dimensions has been generated by randomised interventions, it is common to estimate causal effects with parametric models, sometimes including other covariates to increase power. Such estimates will be potentially biased by the limited expressive capabilities of the parametric representation and corresponding model misspecification issues. If including other covariates, the covariance function representation would have the advantage of including more flexible function spaces that might more accurately reflect the true generative process, with interpretation of the parameters associated with non-randomised covariates being less important. We can therefore expect a more accurate estimation of the underlying causal effect when using a more flexible nonparametric model.

With non-randomised observational data, we may still be interested in achieving some insight into the causal mechanism underlying the data, with all of the appropriate caveats of potential confounding. In this case, the covariance function representation may still help by providing a more flexible function space that increases precision of the causal estimate of interest, and possibly reducing the bias of the estimated causal effect by more accurately modelling the influence of observed confounders.

The separability of the covariance functions necessary to exploit Kronecker structure is potentially an important limitation: it is somewhat analogous to being forced to include interaction effects in a parametric model. If the practitioner is interested in extracting a standalone main effect for the purposes of interpretation, or encoding separate mean functions as random effects at an individual level, then this may hinder interpretation. Further work exploring the interpretation of additively structured Kronecker sub-components would help to elucidate the implications of this constraint. As observed earlier, the stricter constraint of having the training data lie on a (mostly) complete grid is obeyed surprisingly frequently in clinical research, as the importance of performing repeated measurements to isolate different contributions to variation is well-understood.

The use of a Gaussian Process object also opens up the possibility of using Bayesian Optimization\cite{frazier2018tutorial} type algorithms to measure new data points that are optimally informative, according to some acquisition function. The analytical posterior mean and variance of the latent GP object could be used to assess which unsampled areas of the covariate space would be best explored to reduce uncertainty in the effect of interest. Given the relatively strict grid structure necessary,  performing this iteratively may be challenging, but the posterior predictive estimates provided by the fitted GPs could still be used to motivate future research study designs.

In conclusion, in this article, we have demonstrated the ability and utility of scaling Gaussian Process models to large real-world clinical data sets through the use of Kronecker-structure covariance matrices and repeated measurements in the data.


\begin{appendices}

\section{Stan Outputs\label{app:Stan}}

\begin{table}[!h]
\centering
\begin{tabular}{ccccccc}
& mean & sd & l-95\% CI & u-95\% CI & n.eff & Rhat \\
$\rho_1$ & 1.404 & 0.173 & 1.087 & 1.771 & 321.774 & 1.006 \\
$\rho_2$ & 0.97 & 0.076 & 0.823 & 1.12 & 947.797 & 1.007 \\
$\rho_3$ & 1.193 & 0.112 & 0.959 & 1.408 & 202.334 & 1.009 \\
$\alpha_1$ & 0.982 & 0.143 & 0.752 & 1.31 & 623.841 & 1.003 \\
$\alpha_2$ & 1.497 & 0.273 & 1.048 & 2.113 & 304.171 & 1.007 \\
$\alpha_3$ & 1.005 & 0.279 & 0.575 & 1.652 & 1373.194 & 1.005 \\
$\alpha_4$ & 1.735 & 0.42 & 1.068 & 2.718 & 609.914 & 1.006 \\
$\alpha^{(n)}_{1}$ & 0.106 & 0.007 & 0.093 & 0.119 & 1916.695 & 1.002 \\
$\alpha^{(n)}_{2}$ & 0.132 & 0.009 & 0.115 & 0.149 & 650.408 & 1.001 \\
$L^{(n)}_{21}$ & -0.005 & 0.09 & -0.185 & 0.169 & 2174.739 & 1.001 \\
$L^{(n)}_{22}$ & 0.996 & 0.006 & 0.98 & 1 & 2543.158 & 1 \\
$L_{21}$ & -0.135 & 0.143 & -0.416 & 0.149 & 439.501 & 1.006 \\
$L_{22}$ & 0.98 & 0.025 & 0.909 & 1 & 476.384 & 1.007 \\
$L_{31}$ & 0.686 & 0.143 & 0.365 & 0.903 & 2331.114 & 1 \\
$L_{32}$ & -0.08 & 0.196 & -0.432 & 0.319 & 2916.251 & 1 \\
$L_{33}$ & 0.667 & 0.137 & 0.393 & 0.912 & 1995.937 & 1.001 \\
$L_{41}$ & -0.034 & 0.163 & -0.347 & 0.278 & 644.503 & 1.003 \\
$L_{42}$ & 0.867 & 0.085 & 0.646 & 0.976 & 2074.728 & 1 \\
$L_{43}$ & -0.022 & 0.184 & -0.392 & 0.336 & 1443.446 & 1.002 \\
$L_{44}$ & 0.4 & 0.137 & 0.165 & 0.701 & 1730.865 & 1 \\
\end{tabular}
\caption{Posterior summaries of the interpretable parameters of the GP.f model from the first CV fold on the simulated data example from Section \ref{results:simulated}}\label{tab:simulated.GP.f}
\end{table}

\begin{table}[!h]
\centering
\begin{tabular}{ccccccc}
& mean & sd & l-95\% CI & u-95\% CI & n.eff & Rhat \\
$\rho_1$ & 1.833 & 0.171 & 1.507 & 2.188 & 630.434 & 1.004 \\
$\rho_2$ & 1.018 & 0.064 & 0.897 & 1.146 & 1156.583 & 1.004 \\
$\rho_3$ & 24.54 & 32.778 & 5.98 & 89.383 & 1883.495 & 1.003 \\
$\alpha_1$ & 0.746 & 0.095 & 0.579 & 0.954 & 889.352 & 1.003 \\
$\alpha_2$ & 2.076 & 0.261 & 1.613 & 2.624 & 577.406 & 1.007 \\
$\alpha_3$ & 0.863 & 0.264 & 0.48 & 1.487 & 1686.661 & 1.002 \\
$\alpha_4$ & 2.464 & 0.541 & 1.541 & 3.632 & 1534.739 & 1.003 \\
$\alpha^{(n)}_{1}$ & 0.104 & 0.007 & 0.092 & 0.118 & 2983.666 & 1.001 \\
$\alpha^{(n)}_{2}$ & 0.101 & 0.007 & 0.088 & 0.115 & 1881.657 & 1.001 \\
$L^{(n)}_{21}$ & -0.207 & 0.09 & -0.383 & -0.025 & 3693.566 & 1 \\
$L^{(n)}_{22}$ & 0.974 & 0.02 & 0.924 & 0.999 & 3455.251 & 0.999 \\
$L_{21}$ & -0.183 & 0.126 & -0.416 & 0.076 & 364.228 & 1.01 \\
$L_{22}$ & 0.975 & 0.025 & 0.91 & 1 & 411.066 & 1.008 \\
$L_{31}$ & 0.605 & 0.164 & 0.226 & 0.865 & 2594.035 & 1.002 \\
$L_{32}$ & -0.046 & 0.221 & -0.46 & 0.398 & 3936.939 & 1 \\
$L_{33}$ & 0.734 & 0.132 & 0.456 & 0.955 & 2001.942 & 1.003 \\
$L_{41}$ & -0.144 & 0.141 & -0.405 & 0.147 & 599.615 & 1.006 \\
$L_{42}$ & 0.85 & 0.08 & 0.664 & 0.964 & 2309.363 & 1 \\
$L_{43}$ & -0.039 & 0.19 & -0.424 & 0.324 & 1337.541 & 1.008 \\
$L_{44}$ & 0.421 & 0.127 & 0.184 & 0.675 & 2178.473 & 1 \\
$\sigma^{2}_{re}$ & 0.058 & 0.012 & 0.039 & 0.086 & 1102.067 & 1.004 \\
\end{tabular}
\caption{Posterior summaries of the interpretable parameters of the GP.m model from the first CV fold on the simulated data example from Section \ref{results:simulated}}\label{tab:simulated.GP.m}
\end{table}

\begin{table}[!h]
\centering
\begin{tabular}{cccccccc}
& Estimate & sd & l-95\% CI & u-95\% CI & Rhat & Bulk.ESS & Tail.ESS \\
y1.Intercept & 0.002 & 0.068 & -0.134 & 0.135 & 1.001 & 5464.966 & 3092.083 \\
y2.Intercept & -0.161 & 0.055 & -0.268 & -0.055 & 1.002 & 4786.035 & 2940.475 \\
y3.Intercept & 0.058 & 0.147 & -0.234 & 0.345 & 1 & 5457.726 & 3301.1 \\
y4.Intercept & -0.265 & 0.173 & -0.611 & 0.068 & 1 & 5162.581 & 2977.072 \\
y1.x1 & 0.234 & 0.082 & 0.075 & 0.396 & 1.002 & 5932.14 & 3073.043 \\
y1.x2 & 0.421 & 0.073 & 0.277 & 0.565 & 1 & 5867.37 & 2870.77 \\
y1.x3 & -0.03 & 0.078 & -0.185 & 0.121 & 1 & 4867.733 & 2544.01 \\
y2.x1 & -0.49 & 0.067 & -0.623 & -0.362 & 1 & 4672.154 & 2914.568 \\
y2.x2 & 0.446 & 0.059 & 0.335 & 0.559 & 1 & 4995.344 & 2801.483 \\
y2.x3 & 0.359 & 0.062 & 0.24 & 0.477 & 1.001 & 4879.298 & 3101.912 \\
y3.x1 & 0.043 & 0.182 & -0.304 & 0.388 & 1.002 & 5430.894 & 3058.107 \\
y3.x2 & 0.078 & 0.16 & -0.23 & 0.386 & 1.001 & 6038.251 & 3178.676 \\
y3.x3 & -0.032 & 0.169 & -0.36 & 0.301 & 1.001 & 5622.199 & 3016.555 \\
y4.x1 & -0.913 & 0.217 & -1.349 & -0.507 & 1 & 5067.848 & 3007.645 \\
y4.x2 & 0.933 & 0.209 & 0.538 & 1.356 & 1.003 & 4827.501 & 3173.996 \\
y4.x3 & 0.435 & 0.193 & 0.06 & 0.809 & 1 & 4120.251 & 3021.145 \\
\end{tabular}
\caption{Posterior summaries of the parameters of the brms.f model from the first CV fold on the simulated data example from Section \ref{results:simulated}}\label{tab:simulated.brms.f}
\end{table}

\begin{table}[!h]
\centering
\begin{tabular}{cccccccc}
& Estimate & sd & l-95\% CI & u-95\% CI & Rhat & Bulk.ESS & Tail.ESS \\
y1.Intercept & 0.002 & 0.077 & -0.151 & 0.152 & 1.001 & 6402.526 & 2842.214 \\
y2.Intercept & -0.15 & 0.286 & -0.736 & 0.438 & 1.001 & 1398.978 & 1713.599 \\
y3.Intercept & 0.06 & 0.183 & -0.297 & 0.418 & 1.001 & 4684.172 & 2849.119 \\
y4.Intercept & -0.288 & 0.546 & -1.362 & 0.854 & 1.001 & 1799.947 & 2199.357 \\
y1.x1 & 0.235 & 0.082 & 0.07 & 0.391 & 1 & 9351.66 & 2680.745 \\
y1.x2 & 0.423 & 0.073 & 0.28 & 0.565 & 1.002 & 9552.251 & 3006.171 \\
y1.x3 & -0.028 & 0.09 & -0.197 & 0.151 & 1.001 & 5891.755 & 3109.685 \\
y2.x1 & -0.49 & 0.029 & -0.546 & -0.434 & 1 & 8431.978 & 2622.935 \\
y2.x2 & 0.445 & 0.025 & 0.397 & 0.492 & 1.001 & 8605.844 & 2679.603 \\
y2.x3 & 0.35 & 0.324 & -0.333 & 0.996 & 1.002 & 1828.779 & 2024.193 \\
y3.x1 & 0.046 & 0.183 & -0.315 & 0.399 & 1.001 & 7798.346 & 3147.908 \\
y3.x2 & 0.078 & 0.155 & -0.23 & 0.387 & 1.003 & 8106.717 & 2825.376 \\
y3.x3 & -0.03 & 0.219 & -0.451 & 0.391 & 1 & 4265.744 & 2471.228 \\
y4.x1 & -1.154 & 0.257 & -1.671 & -0.668 & 1.001 & 6917.828 & 2769.571 \\
y4.x2 & 1.182 & 0.257 & 0.712 & 1.701 & 1.003 & 8080.056 & 3051.224 \\
y4.x3 & 0.577 & 0.614 & -0.656 & 1.874 & 1.001 & 2099.679 & 2356.275 \\
\end{tabular}
\caption{Posterior summaries of the parameters of the brms.m model from the first CV fold on the simulated data example from Section \ref{results:simulated}}\label{tab:simulated.brms.m}
\end{table}

\begin{table}[!h]
\centering
\begin{tabular}{ccccccc}
& mean & sd & l-95\% CI & u-95\% CI & n.eff & Rhat \\
Side & 0.138 & 0.037 & 0.08 & 0.226 & 1514.041 & 1.002 \\
Time & 0.282 & 0.021 & 0.244 & 0.326 & 479.861 & 1.003 \\
TubeVoltage & 0.298 & 0.099 & 0.151 & 0.539 & 3460.95 & 1 \\
Sex & 1.057 & 0.218 & 0.689 & 1.545 & 2129.349 & 1.004 \\
Age & 0.32 & 0.092 & 0.217 & 0.602 & 75.444 & 1.026 \\
KbyV & 0.976 & 0.253 & 0.338 & 1.391 & 97.965 & 1.017 \\
$\alpha_1$ & 0.584 & 0.049 & 0.495 & 0.686 & 946.77 & 1.007 \\
$\alpha_2$ & 6.708 & 2.442 & 3.748 & 12.746 & 902.727 & 1.008 \\
$\alpha_3$ & 2.963 & 0.593 & 2.029 & 4.351 & 1385.233 & 1.006 \\
$\alpha_4$ & 5.118 & 3.198 & 2.613 & 11.731 & 233.638 & 1.017 \\
$\alpha^{(n)}_{1}$ & 0.447 & 0.017 & 0.415 & 0.481 & 1109.652 & 1.002 \\
$L_{21}$ & 0.767 & 0.072 & 0.607 & 0.884 & 705.094 & 1.002 \\
$L_{22}$ & 0.632 & 0.085 & 0.468 & 0.795 & 684.22 & 1.002 \\
$L_{31}$ & 0.813 & 0.06 & 0.677 & 0.91 & 1004.867 & 1.001 \\
$L_{32}$ & 0.512 & 0.091 & 0.338 & 0.691 & 974.675 & 1.001 \\
$L_{33}$ & 0.249 & 0.053 & 0.154 & 0.361 & 1884.588 & 0.999 \\
$L_{41}$ & 0.776 & 0.076 & 0.604 & 0.903 & 700.609 & 1.004 \\
$L_{42}$ & 0.486 & 0.114 & 0.26 & 0.695 & 693.147 & 1.004 \\
$L_{43}$ & -0.205 & 0.096 & -0.386 & -0.011 & 1136.005 & 1 \\
$L_{44}$ & 0.292 & 0.079 & 0.146 & 0.45 & 1506.603 & 1 \\
\end{tabular}
\caption{Posterior summaries of the interpretable parameters of the GP.f model from the first CV fold on the observational time series example from Section \ref{results:thien}}\label{tab:thien.GP.f}
\end{table}

\begin{table}[!h]
\centering
\begin{tabular}{ccccccc}
& mean & sd & l-95\% CI & u-95\% CI & n.eff & Rhat \\
Side & 0.811 & 0.171 & 0.527 & 1.205 & 1923.84 & 1.001 \\
Time & 0.257 & 0.017 & 0.225 & 0.292 & 507.519 & 1.005 \\
TubeVoltage & 0.249 & 0.097 & 0.117 & 0.48 & 3308.898 & 1 \\
Sex & 1.147 & 0.286 & 0.699 & 1.791 & 3691.531 & 1 \\
Age & 17.824 & 38.394 & 5.412 & 59.024 & 1914.216 & 1.002 \\
KbyV & 6.986 & 3.715 & 3.664 & 14.81 & 1589.906 & 1 \\
$\alpha_1$ & 0.424 & 0.052 & 0.335 & 0.536 & 1307.983 & 1.004 \\
$\alpha_2$ & 9.917 & 6.899 & 3.258 & 29.889 & 1739.195 & 1.002 \\
$\alpha_3$ & 6.743 & 4.765 & 2.485 & 20.149 & 973.293 & 1.001 \\
$\alpha_4$ & 6.568 & 5.372 & 1.985 & 22.38 & 1717.292 & 1.002 \\
$\alpha^{(n)}_{1}$ & 0.334 & 0.016 & 0.303 & 0.366 & 582.19 & 1.003 \\
$L_{21}$ & 0.242 & 0.089 & 0.064 & 0.415 & 2003.83 & 1.002 \\
$L_{22}$ & 0.966 & 0.023 & 0.91 & 0.998 & 2010.449 & 1.001 \\
$L_{31}$ & 0.354 & 0.085 & 0.18 & 0.511 & 1985.88 & 1 \\
$L_{32}$ & 0.848 & 0.045 & 0.748 & 0.926 & 1860.018 & 1.002 \\
$L_{33}$ & 0.377 & 0.065 & 0.256 & 0.514 & 1790.967 & 1.004 \\
$L_{41}$ & 0.211 & 0.099 & 0.016 & 0.402 & 2336.426 & 1 \\
$L_{42}$ & 0.785 & 0.065 & 0.639 & 0.891 & 2637.555 & 1.001 \\
$L_{43}$ & -0.33 & 0.121 & -0.551 & -0.078 & 1495.683 & 1.002 \\
$L_{44}$ & 0.438 & 0.107 & 0.226 & 0.64 & 1183.002 & 1.002 \\
$\sigma^{2}_{re}$ & 0.907 & 0.227 & 0.537 & 1.403 & 1096.364 & 1.001 \\
\end{tabular}
\caption{Posterior summaries of the interpretable parameters of the GP.m model from the first CV fold on the observational time series example from Section \ref{results:thien}}\label{tab:thien.GP.m}
\end{table}

\begin{table}[!h]
\centering
\begin{tabular}{cccccccc}
& Estimate & sd & l-95\% CI & u-95\% CI & Rhat & Bulk.ESS & Tail.ESS \\
HU.Intercept & -1.219 & 0.081 & -1.376 & -1.062 & 1.001 & 7872.147 & 2868.837 \\
Peter.Intercept & -4.675 & 0.483 & -5.666 & -3.759 & 1 & 6087.325 & 2978.142 \\
Sumit.Intercept & -3.235 & 0.354 & -3.939 & -2.575 & 1 & 7279.6 & 3308.639 \\
Thien.Intercept & -9.156 & 0.981 & -11.182 & -7.371 & 1.001 & 3760.612 & 3083.354 \\
HU.Sex & -0.262 & 0.063 & -0.388 & -0.137 & 1 & 6183.659 & 3057.013 \\
HU.Age & -0.085 & 0.027 & -0.137 & -0.031 & 1 & 8696.971 & 2677.263 \\
HU.KbyV & 0.161 & 0.035 & 0.093 & 0.229 & 1.001 & 5213.191 & 3076.073 \\
HU.TubeVoltage & 0 & 0.075 & -0.146 & 0.15 & 1 & 4978.226 & 3690.191 \\
HU.Side & -0.033 & 0.053 & -0.133 & 0.07 & 1.001 & 7692.8 & 2804.257 \\
HU.Time & 2.455 & 0.094 & 2.269 & 2.638 & 1.002 & 9716.016 & 2633.583 \\
Peter.Sex & -1.144 & 0.311 & -1.759 & -0.531 & 1 & 4677.549 & 3330.772 \\
Peter.Age & -0.481 & 0.134 & -0.748 & -0.219 & 1 & 7339.025 & 2992.232 \\
Peter.KbyV & 0.252 & 0.165 & -0.08 & 0.575 & 1 & 5266.446 & 3241.451 \\
Peter.TubeVoltage & 1.184 & 0.364 & 0.48 & 1.906 & 1.001 & 4342.362 & 3319.535 \\
Peter.Side & 0.082 & 0.25 & -0.42 & 0.575 & 1 & 8018.375 & 2380.159 \\
Peter.Time & 9.8 & 0.774 & 8.344 & 11.35 & 1 & 5192.483 & 3105.538 \\
Sumit.Sex & -1.078 & 0.261 & -1.586 & -0.582 & 1 & 5927.317 & 3224.886 \\
Sumit.Age & -0.432 & 0.111 & -0.649 & -0.217 & 1 & 8482.177 & 3230.506 \\
Sumit.KbyV & 0.45 & 0.138 & 0.177 & 0.714 & 1.001 & 5359.511 & 3409.304 \\
Sumit.TubeVoltage & 0.851 & 0.306 & 0.258 & 1.445 & 1 & 4939.906 & 3211.057 \\
Sumit.Side & -0.022 & 0.213 & -0.429 & 0.388 & 1.001 & 8566.185 & 2964.559 \\
Sumit.Time & 6.276 & 0.485 & 5.367 & 7.223 & 1.001 & 7182.14 & 3646.269 \\
Thien.Sex & -0.817 & 0.451 & -1.704 & 0.059 & 1 & 5344.679 & 3392.033 \\
Thien.Age & -0.464 & 0.187 & -0.831 & -0.107 & 1 & 7095.642 & 3507.823 \\
Thien.KbyV & -0.02 & 0.238 & -0.489 & 0.443 & 1.002 & 5797.254 & 3370.95 \\
Thien.TubeVoltage & 1.695 & 0.542 & 0.693 & 2.787 & 1 & 4477.558 & 3326.391 \\
Thien.Side & 0.156 & 0.361 & -0.544 & 0.882 & 1 & 6439.037 & 2641.699 \\
Thien.Time & 20.262 & 1.988 & 16.651 & 24.36 & 1 & 3534.26 & 2875.124 \\
\end{tabular}
\caption{Posterior summaries of the parameters of the brms.f model from the first CV fold on the observational time series example from Section \ref{results:thien}}\label{tab:thien.brms.f}
\end{table}

\begin{table}[!h]
\centering
\begin{tabular}{cccccccc}
& Estimate & sd & l-95\% CI & u-95\% CI & Rhat & Bulk.ESS & Tail.ESS \\
HU.Intercept & -1.229 & 0.11 & -1.44 & -1.005 & 1.001 & 1883.872 & 2635.57 \\
Peter.Intercept & -6.888 & 0.929 & -8.859 & -5.181 & 1 & 2155.894 & 2448.99 \\
Sumit.Intercept & -4.592 & 0.74 & -6.093 & -3.181 & 1.003 & 2059.998 & 2829.259 \\
Thien.Intercept & -12.612 & 1.799 & -16.488 & -9.523 & 1.003 & 1913.781 & 2247.858 \\
HU.Sex & -0.261 & 0.107 & -0.471 & -0.053 & 1 & 1326.375 & 2245.068 \\
HU.Age & -0.083 & 0.045 & -0.173 & 0.004 & 1.003 & 1412.836 & 2148.39 \\
HU.KbyV & 0.159 & 0.055 & 0.049 & 0.264 & 1.003 & 1510.37 & 2118.139 \\
HU.TubeVoltage & 0.006 & 0.121 & -0.233 & 0.238 & 1.002 & 1436.848 & 2183.957 \\
HU.Side & -0.03 & 0.049 & -0.125 & 0.066 & 1 & 5286.117 & 3292.53 \\
HU.Time & 2.45 & 0.086 & 2.281 & 2.617 & 1.001 & 5186.381 & 2701.354 \\
Peter.Sex & -1.657 & 0.802 & -3.306 & -0.14 & 1.001 & 1390.627 & 2088.921 \\
Peter.Age & -0.688 & 0.332 & -1.355 & -0.03 & 1.003 & 1265.427 & 1992.546 \\
Peter.KbyV & 0.303 & 0.395 & -0.459 & 1.077 & 1.002 & 1455.29 & 2067.752 \\
Peter.TubeVoltage & 1.715 & 0.882 & -0.029 & 3.455 & 1.003 & 1361.107 & 2431.833 \\
Peter.Side & 0.146 & 0.304 & -0.452 & 0.747 & 1.003 & 5842.318 & 2887.942 \\
Peter.Time & 14.326 & 1.286 & 11.963 & 17.022 & 1 & 2992.62 & 2513.319 \\
Sumit.Sex & -1.457 & 0.71 & -2.848 & -0.063 & 1 & 1349.48 & 2298.628 \\
Sumit.Age & -0.586 & 0.291 & -1.164 & -0.026 & 1.002 & 1313.955 & 2027.691 \\
Sumit.KbyV & 0.585 & 0.353 & -0.133 & 1.274 & 1.003 & 1336.028 & 1903.154 \\
Sumit.TubeVoltage & 1.132 & 0.788 & -0.466 & 2.695 & 1.001 & 1316.225 & 1989.443 \\
Sumit.Side & -0.021 & 0.253 & -0.521 & 0.481 & 1.001 & 5200.054 & 2922.916 \\
Sumit.Time & 8.754 & 0.713 & 7.429 & 10.248 & 1.003 & 3797.535 & 3110.568 \\
Thien.Sex & -1.15 & 0.857 & -2.879 & 0.547 & 1 & 1805.37 & 2631.778 \\
Thien.Age & -0.64 & 0.366 & -1.367 & 0.049 & 1.001 & 1739.675 & 2030.872 \\
Thien.KbyV & -0.1 & 0.44 & -0.981 & 0.751 & 1.002 & 1957.174 & 2417.188 \\
Thien.TubeVoltage & 2.423 & 1.008 & 0.498 & 4.525 & 1.002 & 1914.17 & 2449.821 \\
Thien.Side & 0.281 & 0.42 & -0.521 & 1.119 & 1.001 & 5454.051 & 2826.627 \\
Thien.Time & 27.753 & 3.479 & 21.682 & 35.474 & 1.003 & 1881.15 & 2370.816 \\
\end{tabular}
\caption{Posterior summaries of the parameters of the brms.m model from the first CV fold on the observational time series example from Section \ref{results:thien}}\label{tab:thien.brms.m}
\end{table}

\begin{table}[!h]
\centering
\begin{tabular}{ccccccc}
& mean & sd & l-95\% CI & u-95\% CI & n.eff & Rhat \\
Location.Iliac & 1.286 & 0.074 & 1.146 & 1.439 & 1513.542 & 1.003 \\
Location.Femoral & 9.77 & 1.739 & 7.113 & 14.038 & 445.735 & 1.005 \\
Location.Popliteal & 14.298 & 4.277 & 8.887 & 23.962 & 561.439 & 1.003 \\
keV.40 & 0.237 & 0.026 & 0.187 & 0.29 & 759.765 & 1.002 \\
keV.50 & 0.113 & 0.014 & 0.088 & 0.142 & 744.719 & 1.002 \\
Randomisation & 0.292 & 0.038 & 0.228 & 0.37 & 633.115 & 1.009 \\
Sex & 1.052 & 0.078 & 0.905 & 1.212 & 1591.639 & 1.007 \\
Age & 0.237 & 0.015 & 0.202 & 0.265 & 44.326 & 1.092 \\
Flow & 0.348 & 0.03 & 0.305 & 0.439 & 38.1 & 1.116 \\
$\alpha_1$ & 1.012 & 0.054 & 0.914 & 1.128 & 480.564 & 1.003 \\
$\alpha_2$ & 0.619 & 0.035 & 0.555 & 0.691 & 758.82 & 1.001 \\
$\alpha_3$ & 4.936 & 0.776 & 3.694 & 6.722 & 1065.099 & 1.002 \\
$\alpha_4$ & 15.009 & 2.419 & 11.154 & 20.61 & 1338.5 & 1 \\
$\alpha^{(n)}_{1}$ & 0.294 & 0.004 & 0.286 & 0.303 & 1306.745 & 1.003 \\
$\alpha^{(n)}_{2}$ & 0.715 & 0.009 & 0.697 & 0.733 & 1576.561 & 1.001 \\
$L^{(n)}_{21}$ & -0.144 & 0.019 & -0.181 & -0.107 & 2935.456 & 1.001 \\
$L^{(n)}_{22}$ & 0.989 & 0.003 & 0.984 & 0.994 & 2885.614 & 1.001 \\
$L_{21}$ & 0.732 & 0.033 & 0.661 & 0.79 & 694.7 & 1.003 \\
$L_{22}$ & 0.68 & 0.035 & 0.613 & 0.75 & 700.848 & 1.003 \\
$L_{31}$ & 0.602 & 0.044 & 0.512 & 0.685 & 1118.582 & 1.003 \\
$L_{32}$ & -0.032 & 0.073 & -0.176 & 0.114 & 905.705 & 1.007 \\
$L_{33}$ & 0.793 & 0.034 & 0.722 & 0.857 & 1132.266 & 1.002 \\
$L_{41}$ & 0.71 & 0.031 & 0.647 & 0.767 & 985.695 & 1.003 \\
$L_{42}$ & -0.015 & 0.061 & -0.135 & 0.104 & 492.138 & 1.006 \\
$L_{43}$ & 0.321 & 0.057 & 0.205 & 0.429 & 511.796 & 1.012 \\
$L_{44}$ & 0.619 & 0.038 & 0.545 & 0.693 & 628.993 & 1.005 \\
\end{tabular}
\caption{Posterior summaries of the interpretable parameters of the GP.f model from the first CV fold on the RCT example from Section \ref{results:helgestad}}\label{tab:helgestad.GP.f}
\end{table}

\begin{table}[!h]
\centering
\begin{tabular}{ccccccc}
& mean & sd & l-95\% CI & u-95\% CI & n.eff & Rhat \\
Location.Iliac & 1.071 & 0.095 & 0.9 & 1.275 & 496.725 & 1.008 \\
Location.Femoral & 1.69 & 0.112 & 1.482 & 1.931 & 718.188 & 1.002 \\
Location.Popliteal & 1.776 & 0.125 & 1.552 & 2.039 & 640.221 & 1.005 \\
keV.40 & 0.102 & 0.012 & 0.079 & 0.127 & 1185.38 & 1.002 \\
keV.50 & 0.054 & 0.007 & 0.041 & 0.068 & 1123.026 & 1 \\
Randomisation & 0.12 & 0.026 & 0.078 & 0.18 & 1160.001 & 1 \\
Sex & 1.31 & 0.166 & 1.022 & 1.668 & 2934.555 & 1 \\
Age & 35.694 & 12.951 & 19.5 & 69.44 & 812.477 & 1.004 \\
Flow & 11.488 & 2.314 & 7.528 & 16.652 & 471.008 & 1.001 \\
$\alpha_1$ & 1 & 0.077 & 0.861 & 1.16 & 691.075 & 1.006 \\
$\alpha_2$ & 1.317 & 0.113 & 1.11 & 1.548 & 632.407 & 1.006 \\
$\alpha_3$ & 8.796 & 1.699 & 6.114 & 12.616 & 909.123 & 1.005 \\
$\alpha_4$ & 14.696 & 3.03 & 10.203 & 21.863 & 877.625 & 1.005 \\
$\alpha^{(n)}_{1}$ & 0.243 & 0.003 & 0.237 & 0.249 & 2819.81 & 1 \\
$\alpha^{(n)}_{2}$ & 0.567 & 0.008 & 0.552 & 0.582 & 1474.628 & 1 \\
$L^{(n)}_{21}$ & -0.12 & 0.018 & -0.158 & -0.085 & 2945.145 & 1 \\
$L^{(n)}_{22}$ & 0.993 & 0.002 & 0.988 & 0.996 & 2900.402 & 1 \\
$L_{21}$ & 0.295 & 0.041 & 0.217 & 0.374 & 770.558 & 1.004 \\
$L_{22}$ & 0.955 & 0.013 & 0.927 & 0.976 & 768.216 & 1.004 \\
$L_{31}$ & 0.46 & 0.046 & 0.366 & 0.546 & 1245.77 & 1.002 \\
$L_{32}$ & -0.282 & 0.071 & -0.416 & -0.137 & 1191.492 & 1.003 \\
$L_{33}$ & 0.837 & 0.033 & 0.771 & 0.898 & 1171.472 & 1.003 \\
$L_{41}$ & 0.377 & 0.042 & 0.292 & 0.458 & 978.818 & 1.001 \\
$L_{42}$ & -0.053 & 0.061 & -0.169 & 0.067 & 932.93 & 1.005 \\
$L_{43}$ & 0.106 & 0.079 & -0.046 & 0.262 & 277.257 & 1.011 \\
$L_{44}$ & 0.912 & 0.02 & 0.871 & 0.948 & 907.221 & 1.003 \\
$\sigma^{2}_{re}$ & 0.121 & 0.018 & 0.09 & 0.159 & 769.687 & 1.003 \\
\end{tabular}
\caption{Posterior summaries of the interpretable parameters of the GP.m model from the first CV fold on the RCT example from Section \ref{results:helgestad}}\label{tab:helgestad.GP.m}
\end{table}

\begin{table}[!h]
\centering
\begin{tabular}{cccccccc}
& Estimate & sd & l-95\% CI & u-95\% CI & Rhat & Bulk.ESS & Tail.ESS \\
HU.Intercept & 0.275 & 0.04 & 0.196 & 0.357 & 1 & 10453.102 & 2692.506 \\
SD.Intercept & -0.368 & 0.048 & -0.462 & -0.273 & 1.002 & 8602.66 & 3274.616 \\
Sumit.Intercept & 2.844 & 0.224 & 2.408 & 3.296 & 1.001 & 7851.73 & 3100.73 \\
DanLevi.Intercept & 2.046 & 0.158 & 1.735 & 2.343 & 1 & 7434.289 & 2993.31 \\
HU.Sex & -0.043 & 0.029 & -0.1 & 0.014 & 0.999 & 7290.619 & 3315.779 \\
HU.Age & 0.076 & 0.013 & 0.051 & 0.101 & 1.001 & 8695.496 & 2708.414 \\
HU.Flowmls & 0.149 & 0.015 & 0.119 & 0.178 & 1 & 7565.725 & 3049.803 \\
HU.Side & 0 & 0.027 & -0.05 & 0.054 & 1.001 & 8424.888 & 3120.189 \\
HU.Location & -0.085 & 0.013 & -0.11 & -0.059 & 1.001 & 9835.406 & 2940.605 \\
HU.keV & -0.625 & 0.013 & -0.65 & -0.6 & 1.005 & 9179.524 & 2829.488 \\
SD.Sex & -0.272 & 0.034 & -0.339 & -0.204 & 1.001 & 7248.65 & 3262.452 \\
SD.Age & -0.021 & 0.015 & -0.051 & 0.009 & 1 & 8169.73 & 3188.157 \\
SD.Flowmls & 0.226 & 0.018 & 0.192 & 0.261 & 1.003 & 7449.155 & 3557.409 \\
SD.Side & 0.091 & 0.032 & 0.028 & 0.156 & 1.001 & 8420.121 & 3258.472 \\
SD.Location & 0.183 & 0.015 & 0.153 & 0.213 & 1.001 & 9028.951 & 2844.361 \\
SD.keV & -0.271 & 0.015 & -0.301 & -0.241 & 1.001 & 10044.46 & 2975.49 \\
Sumit.Sex & 0.053 & 0.16 & -0.268 & 0.357 & 1 & 7327.927 & 3328.309 \\
Sumit.Age & -0.088 & 0.073 & -0.234 & 0.049 & 1 & 10120.038 & 3351.897 \\
Sumit.Flowmls & 0.067 & 0.081 & -0.093 & 0.22 & 1 & 6514.66 & 3419.375 \\
Sumit.Side & 0.139 & 0.148 & -0.152 & 0.432 & 1.002 & 8806.937 & 2611.821 \\
Sumit.Location & 0.006 & 0.068 & -0.127 & 0.141 & 1.002 & 8812.139 & 2832.606 \\
Sumit.keV & -0.847 & 0.069 & -0.988 & -0.717 & 1.001 & 6880.567 & 3395.213 \\
DanLevi.Sex & 0.204 & 0.115 & -0.024 & 0.43 & 1.001 & 7937.028 & 2831.09 \\
DanLevi.Age & 0.02 & 0.05 & -0.078 & 0.115 & 1.002 & 9383.669 & 2899.218 \\
DanLevi.Flowmls & 0.194 & 0.057 & 0.083 & 0.308 & 1 & 7261.622 & 3446.642 \\
DanLevi.Side & 0.042 & 0.101 & -0.153 & 0.244 & 1 & 8716.294 & 3142.675 \\
DanLevi.Location & -0.141 & 0.047 & -0.231 & -0.05 & 1.002 & 7775.343 & 2986.236 \\
DanLevi.keV & -1.296 & 0.05 & -1.396 & -1.197 & 1 & 6444.788 & 3221.858 \\
\end{tabular}
\caption{Posterior summaries of the parameters of the brms.f model from the first CV fold on the RCT example from Section \ref{results:helgestad}}\label{tab:helgestad.brms.f}
\end{table}

\begin{table}[!h]
\centering
\begin{tabular}{cccccccc}
& Estimate & sd & l-95\% CI & u-95\% CI & Rhat & Bulk.ESS & Tail.ESS \\
HU.Intercept & 0.238 & 0.088 & 0.062 & 0.401 & 1.026 & 187.541 & 469.566 \\
SD.Intercept & -0.386 & 0.075 & -0.535 & -0.24 & 1.005 & 418.06 & 1129.855 \\
Sumit.Intercept & 5.513 & 0.645 & 4.332 & 6.825 & 1.006 & 724.949 & 1647.996 \\
DanLevi.Intercept & 5.439 & 0.802 & 3.937 & 7.059 & 1.009 & 402.24 & 958.231 \\
HU.Sex & -0.024 & 0.121 & -0.247 & 0.218 & 1.028 & 173.245 & 268.353 \\
HU.Age & 0.078 & 0.049 & -0.017 & 0.174 & 1.019 & 255.92 & 531.007 \\
HU.Flowmls & 0.152 & 0.059 & 0.039 & 0.267 & 1.016 & 210.331 & 380.7 \\
HU.Side & 0.004 & 0.014 & -0.024 & 0.032 & 1 & 5957.802 & 2980.97 \\
HU.Location & -0.075 & 0.007 & -0.088 & -0.062 & 1.002 & 6749.447 & 2902.081 \\
HU.keV & -0.626 & 0.007 & -0.639 & -0.612 & 1.001 & 8216.338 & 2970.541 \\
SD.Sex & -0.268 & 0.092 & -0.438 & -0.084 & 1.009 & 305.963 & 568.382 \\
SD.Age & -0.024 & 0.037 & -0.097 & 0.048 & 1.011 & 408.812 & 947.47 \\
SD.Flowmls & 0.222 & 0.044 & 0.135 & 0.308 & 1.006 & 347.06 & 930.257 \\
SD.Side & 0.089 & 0.029 & 0.032 & 0.145 & 1.002 & 5594.598 & 2939.804 \\
SD.Location & 0.19 & 0.013 & 0.164 & 0.216 & 1.001 & 6120.002 & 3129.237 \\
SD.keV & -0.272 & 0.013 & -0.297 & -0.246 & 1.001 & 6320.706 & 2932.901 \\
Sumit.Sex & 0.267 & 0.698 & -1.066 & 1.655 & 1.008 & 347.219 & 766.086 \\
Sumit.Age & -0.128 & 0.313 & -0.771 & 0.46 & 1.012 & 502.922 & 1384.839 \\
Sumit.Flowmls & -0.034 & 0.356 & -0.761 & 0.64 & 1.004 & 405.3 & 786.352 \\
Sumit.Side & 0.25 & 0.193 & -0.132 & 0.636 & 1.002 & 4853.972 & 2810.804 \\
Sumit.Location & 0.006 & 0.085 & -0.158 & 0.172 & 1.001 & 5430.007 & 3161.704 \\
Sumit.keV & -1.257 & 0.095 & -1.451 & -1.077 & 1.001 & 5131.408 & 3045.637 \\
DanLevi.Sex & 0.803 & 0.974 & -1.141 & 2.675 & 1.008 & 305.69 & 595.933 \\
DanLevi.Age & -0.169 & 0.426 & -1.004 & 0.67 & 1.007 & 462.015 & 1039.884 \\
DanLevi.Flowmls & 0.285 & 0.496 & -0.7 & 1.238 & 1.002 & 410.709 & 1011.654 \\
DanLevi.Side & 0.14 & 0.19 & -0.231 & 0.503 & 1 & 5126.812 & 3077.509 \\
DanLevi.Location & -0.39 & 0.091 & -0.565 & -0.21 & 1 & 4979.411 & 3335.896 \\
DanLevi.keV & -3.116 & 0.136 & -3.389 & -2.86 & 1 & 3505.023 & 2501.601 \\
\end{tabular}
\caption{Posterior summaries of the parameters of the brms.m model from the first CV fold on the RCT example from Section \ref{results:helgestad}}\label{tab:helgestad.brms.m}
\end{table}

\newpage
\clearpage
\section{Random effects in the Intrinsic Coregionalization Model}\label{app:re-icm}
In this appendix, we demonstrate the effect of adding a random effect to an Intrinsic Coregionalization Model (ICM), in the simplest case where the covariance functions over the inputs and the outputs are both linear. Recall that the ICM for a multi-output GP over $m$ outputs can be written as 
\begin{align}
    \begin{bmatrix}
        f_1 \\
        \vdots \\
        f_m 
    \end{bmatrix} \sim \mathcal{GP}(0,B \  \kappa_x(x,x')),
\end{align}
where $B$ is the $m \times m$ coregionalisation matrix giving the covariance over the outputs, and $\kappa_x(x,x')$ the covariance function over the inputs. In the case of a fully observed dataset, we can write this using the Kronecker product $\mathbf{f}(\mathbf{X})\sim\mathcal{N}(0,B\otimes K_x)$. By exploiting the connections between Gaussian Processes and Bayesian linear regression, in the case of a linear kernel over the inputs, $\kappa_x(\mathbf{x},\mathbf{x}')=1+\mathbf{x}^T\mathbf{x}'$, the ICM is equivalent to the following linear model:
\begin{align}\label{eq:ICMequivalence}
    \begin{bmatrix}
        \mathbf{f}_1(\mathbf{X}) \\
        \vdots \\
        \mathbf{f}_m(\mathbf{X})
    \end{bmatrix}=
    \begin{bmatrix}
        \mathbf{X} & 0 & \cdots & 0 \\
        0 & \mathbf{X} & \cdots & 0 \\
        \vdots & \vdots & \ddots & \vdots \\
        0 & 0 & \cdots & \mathbf{X}
    \end{bmatrix}
    \begin{bmatrix}
        \boldsymbol{\beta}_1 \\
        \boldsymbol{\beta}_2 \\
        \vdots \\
        \boldsymbol{\beta_m}
    \end{bmatrix}, \ \ 
       \begin{bmatrix}
        \boldsymbol{\beta}_1 \\
        \boldsymbol{\beta}_2 \\
        \vdots \\
        \boldsymbol{\beta_m}
    \end{bmatrix} \sim \mathcal{N}(0,B\otimes \mathcal{I}_{(p+1)}),
\end{align}
where $\mathbf{X}\in\mathbb{R}^{m\times (p+1)}$ denotes the design matrix of a linear regression including the intercept, $\boldsymbol{\beta_j}\in\mathbb{R^{(p+1)}}$ the corresponding output-specific coefficient vector, and $\mathcal{I}_p$ the $(p+1)$-dimensional identity matrix. Hence, this is equivalent to fitting a linear model to each output, with output-specific coefficients $\boldsymbol{\beta}_j$ for $j=1,\ldots,m$. Due to the joint distribution over the coefficient vectors, these linear models are correlated in the prior, allowing the models to influence each other, borrowing strength across outputs.

When working with multiple covariance functions, that are multiplied together, we need to consider the effect on the induced feature space. Consider a set of covariates that can be blocked into two sets $\mathbf{X}=[\mathbf{X}^{(1)},\mathbf{X}^{(2)}]$, where $\mathbf{X}^{(2)}$ correspond to patient-specific covariates, while $\mathbf{X}^{(1)}$ are other covariates in the model. In this paper, we consider covariance structures that decompose across these groups of covariates, e.g. $\kappa(\mathbf{X},\mathbf{X}')=\kappa_1(\mathbf{X}^{(1)},\mathbf{X}^{(1)})\kappa_2(\mathbf{X}^{(2)},\mathbf{X}^{(2)})$. Generally, multiplying together kernels has the effect of modelling interactions between the covariates, e.g. if both $\kappa_1$ and $\kappa_2$ are linear kernels, the induced features space is $\phi_{1\times 2}=\{1,x^{(1)}_1,\ldots,x^{(1)}_{p_1},x^{(2)}_1,\ldots,x^{(2)}_{p_2},x^{(1)}_1x^{(2)}_1,\ldots,x^{(1)}_{p_1}x^{(2)}_{p_2}\}$, and GP regression is equivalent to performing a linear regression using the extended basis $\phi_{1\times 2}$. Similarly in the multi-output setting GP regression becomes equivalent to the model in equation (\ref{eq:ICMequivalence}), but with the $m \times ((p_1+1)(p_2+1))$ matrix $\Phi_{1 \times 2}$ taking the place of $X$, and modulo some dimensionality changes on the identity matrix in the prior over the coefficients.

Further complicating the model by adding a random effect to $\kappa_2$ as in Section \ref{seq:randomeffects}:
\begin{align}
    \begin{bmatrix}
        f_1 \\
        \vdots \\
        f_m 
    \end{bmatrix} \sim \mathcal{GP}\left(0,B \ \kappa_1\left(\mathbf{X}^{(1)},\mathbf{X}^{(1)}\right)\left(\kappa_2\left(\mathbf{X}^{(2)},\mathbf{X}^{(2)}\right)+\gamma^2 \mathcal{I}_{m}\right)\right),
\end{align}
has the effect of adding extra variance to the coefficients corresponding to the non-patient specific covariates. The model becomes
\begin{align}
    \begin{bmatrix}
        \mathbf{f}_1(\mathbf{X}) \\
        \vdots \\
        \mathbf{f}_m(\mathbf{X})
    \end{bmatrix}=
    \begin{bmatrix}
       \Phi_{1\times 2} & 0 & \cdots & 0 \\
        0 & \Phi_{1\times 2} & \cdots & 0 \\
        \vdots & \vdots & \ddots & \vdots \\
        0 & 0 & \cdots & \Phi_{1\times 2}
    \end{bmatrix}
    \begin{bmatrix}
        \boldsymbol{\beta}_1 \\
        \boldsymbol{\beta}_2 \\
        \vdots \\
        \boldsymbol{\beta_m}
    \end{bmatrix}, \ \ \ \text{where} \ \ \
       \begin{bmatrix}
        \boldsymbol{\beta}_1 \\
        \boldsymbol{\beta}_2 \\
        \vdots \\
        \boldsymbol{\beta_m}
    \end{bmatrix} \sim \mathcal{N}\left(0,B\otimes \tilde{\mathcal{I}}_{(p_1+1)(p_2+1)}\right),
\end{align}
where $\tilde{\mathcal{I}}_{(p_1+1)(p_2+1)}$ is a block diagonal matrix with blocks $\{(1+\gamma^2)\mathcal{I}_{(p_1+1)},\mathcal{I}_{p_2},\mathcal{I}_{p_1p_2}\}$.

\end{appendices}

\section{Competing interests}
The authors declare no potential conflict of interests.

\section{Author contributions statement}

OT helped with conceptualising the project, constructing the code, running the experiments, and writing the manuscript. LR helped with model design, providing the starting codebase, writing the manuscript, and deriving theoretical results.

\section{Acknowledgments}

We thank Thien Trung Tran, Cathrine Helgestad Kristiansen and Peter Lauritzen for providing the radiological data used in this article.

\bibliographystyle{plain}
\bibliography{oup-authoring-template}


\begin{biography}{}{\author{Owen Thomas.} Owen Thomas is a Senior Researcher at the Health Services Research Unit (HØKH) at Akershus University Hospital in Lørenskog, Norway. He previously completed his DPhil at the Department of Statistics of the University of Oxford, UK, and a postdoctoral fellowship at the Oslo Centre for Biostatistics and Epidemiology (OCBE), of the University of Oslo, Norway.}
\end{biography}

\begin{biography}{}{\author{Leiv Rønneberg.} Leiv Rønneberg is a Research Associate at the MRC Biostatistics Unit at the University of Cambridge, Cambridge UK. He completed his PhD at the Oslo Centre for Biostatistics and Epidemiology (OCBE), at the University of Oslo, Norway.}
\end{biography}

\end{document}